\documentclass[epj]{svjour}
\usepackage{graphics}
\usepackage{amsmath,amssymb,color}
\begin{document}

\newcommand{\red}{\textcolor{red}}
\newcommand{\blue}{\textcolor{blue}}

\title{Magnetic properties of the quantum spin-$\frac{1}{2}$ $XX$ diamond chain: \\
The Jordan-Wigner approach}
\author{Taras Verkholyak\inst{1,2},
        Jozef Stre\v{c}ka\inst{2},
        Michal Ja\v{s}\v{c}ur\inst{2},
        Johannes Richter\inst{3}}
%
%
%
\authorrunning{T. Verkholyak et al.}
\institute{Institute for Condensed Matter Physics,
             National Academy of Sciences of Ukraine,
             1 Svientsitskii Street, L'viv-11, 79011, Ukraine
             \and
             Department of Theoretical Physics and Astrophysics,
             Institute of Physics, P. J. \v{S}af\'{a}rik University,
             Park Angelinum 9, 040 01 Ko\v{s}ice, Slovak Republic
             \and
             Institut f\"ur Theoretische Physik,
             Otto-von-Guericke-Universit\"at Magdeburg,
             P.O. Box 4120, 39016 Magdeburg, Germany }
\date{Received: date / Revised version: date}
%
\abstract{
The Jordan-Wigner transformation is applied to study magnetic properties of
the quantum spin-$\frac{1}{2}$ $XX$ model on the diamond chain. Generally,
the Hamiltonian of this quantum spin system can be represented in terms of
spinless fermions in the presence of a gauge field and different gauge-invariant
ways of assigning the spin-fermion transformation are considered.
Additionally, we analyze
general properties of a free-fermion chain, where all gauge terms are neglected
and discuss their relevance for the quantum spin system. A consideration of interaction
terms in the fermionic Hamiltonian rests upon the Hartree-Fock procedure after
fixing the appropriate gauge. Finally, we discuss the magnetic properties
of this quantum spin model at zero as well as non-zero temperatures and analyze
the validity of the approximation used through a comparison with the results
of the exact diagonalization method for finite (up to 36 spins) chains.
Besides the $m=1/3$ plateau the most prominent feature of the
magnetization curve is a jump at intermediate field present for certain
values of the frustrating vertical bond.
}



\maketitle


\section{Introduction}
\label{intro}
Low-dimensional quantum spin models on frustrated lattices represent
objects of intense current research
(see e.g. Refs.\cite{misguich,mikeska,richter,derzhko2007} for the recent reviews).
The analytical study of such models is however quite involved,
since the interplay of competing interactions, quantum fluctuations
and magnetic field may produce at sufficiently low temperatures
a diversity of unusual quantum phases. A dimerization in the ground state,
localized excitations, magnetization jumps and plateaus are the most typical phenomena,
which might possibly appear in geometrically frustrated quantum spin
models\cite{misguich,mikeska,richter,derzhko2007,prl02,honecker04}.

The spin-$\frac{1}{2}$ Heisenberg model on the diamond chain is an example of a frustrated spin system,
which represents one of the simplest quantum models with the exactly known monomer-dimer ground state.
When the nearest-neighbor spins from a diamond chain are coupled through antiferromagnetic
interactions, this model actually exhibits either ferrimagnetically ordered ground
states or the
disordered tetramer-dimer and dimer-monomer spin-fluid phases depending on
the relative strength
of the geometric frustration \cite{tks}.
A number of  exact results for low-temperature properties are also available
at sufficiently high magnetic fields close to a saturation value
at which this quantum system shows a magnetization jump towards the completely polarized phase
                 (see e.g. Refs.~\cite{derzhko2007,prl02} and
\cite{honecker04}).
Exact results can be also found for the special Ising-Heisenberg diamond chain when the
only quantum interaction is between spins on vertical bonds and one may apply
the decoration-iteration transformation to study this simplified model rigorously in the whole temperature range \cite{canova}.
Most recently the ground states of the mixed diamond chain with higher spins
have been found rigorously
\cite{hida1}.
Quantum phase transitions and finite-temperature properties of the
diamond chain with the mixed spins 1 and 1/2 have been studied in Refs.\cite{hida2,hida3}.

The ground-state phase diagram of the spin-$\frac{1}{2}$ Heisenberg model on the distorted diamond chain
cannot be found exactly in general, but the extensive study of the ground phase has been performed
in Refs.~\cite{okamoto1999,okamoto2003} by means of the exact diagonalization and some perturbative approaches.
The effect of the exchange anisotropy on the ground state properties of the spin-$\frac{1}{2}$
$XXZ$ diamond chain was considered in Refs.~\cite{okamoto2005,okamoto2005-2,okamoto2007},
where the interesting inversion
phenomenon has been theoretically predicted. In the case of the easy-plane anisotropy,
the ground-state phase diagram contains an additional N\'eel phase as a result of the
interplay between the exchange anisotropy, geometric frustration and quantum fluctuations.

The experimental detection of the frustrated diamond chain in azurite Cu$_3$(CO$_3$)$_2$(OH)$_2$ \cite{kikuchi},
the ferrimagnetic diamond chain in organic-radical system \cite{hosokoshi},
as well as, the ferromagnetic diamond chains in polymeric coordination compounds Bi$_4$Cu$_3$V$_2$O$_{14}$ \cite{sakurai},
Cu$_3$(TeO$_3$)$_2$Br$_2$ \cite{uematsu} and Cu$_3$(N$_3$)$_6$(DMF)$_2$ \cite{lazari}
have stimulated a number of experimental
\cite{ohta2003,kikuchi2004,ohta2004,kikuchi2005a,kikuchi2005b,rule}
and theoretical \cite{canova,mikeska2008,fu,gu,azurite2010} studies over the last few years.
The study of the
field-dependent magnetization curve in the ground state, dynamic properties
and the determination of spin couplings are the most discussed problems nowadays
\cite{rule,mikeska2008,gu,azurite2010}.

Most of the previous theoretical treatments are based on numerical
techniques. Hence an analytical approach covering the ground state as well
as thermodynamic properties is desirable.
A promising method is the Jordan-Wigner fermionization of the spin degrees of
freedom. Such an approach has been presented in Refs. \cite{fu,gu}. However, the
results presented there suffer from the neglecting of phase factors and,
therefore, they cannot be considered as reliable (see our discussion in
Sect.~\ref{free}).
Note that the Jordan-Wigner transformation also gives a useful representation
for the quantum models on two-leg ladders \cite{azzouz1994,nunner,verkholyak2006}.
In the case of the railroad ladder
approximative considerations allow to describe not only the ground state,
but also the dynamic properties \cite{nunner}.

The goal of the present paper is to describe thermodynamic and
magnetic properties of the quantum spin-$\frac{1}{2}$ $XX$ diamond chain
in fermionic language by means of the Jordan-Wigner transformation
and to complement our findings by exact diagonalization data.
It should be stressed,
however, that the diamond chain model is not reduced after
performing the Jordan-Wigner transformation to free fermions.
The $XX$ interaction terms are not mapped to the
two-fermion terms but they might contain phase factors whose
specific form will basically depend on a particular choice of the
nonlocal Jordan-Wigner transformation
(see the discussion on the railroad ladder in Ref.~\cite{nunner}).
On the other
hand, all different fermionic representations are connected
through appropriate gauge transformations. Because of the phase
factors, which effectively introduce interactions between fermions
and act as operators, the quantum diamond chain model cannot be
treated rigorously anymore and some further approximations are
required.
Note that this kind of fermionic
representation is extremely useful for studying dynamic properties
of quantum spin-$\frac{1}{2}$ models (see e.g. Refs.~\cite{mueller1981,derzhko2005}).

The outline of this paper is as follows. In Sec.~\ref{jordan-wigner},
the Jordan-Wigner transformation for the
spin-$\frac{1}{2}$ $XX$ model on the diamond chain will be considered. The free-fermion model
and its relation to the quantum spin model is analyzed in Sec.~\ref{free}.
The Hartree-Fock approximation
is applied to the fermionic counterpart of the symmetric and distorted diamond chain
in Secs.~\ref{distorted_chain_hf} and \ref{symmetric_chain_hf}.
The ground-state properties will be finally calculated and compared
with the known exact and numerical results there.
The thermodynamic properties will be considered in Sec.~\ref{temperature_effect}
and the obtained results are summarized in Sec.~\ref{conclusions}.

\section{Jordan-Wigner transformation for the spin-$\frac{1}{2}$ $XX$ diamond chain}
\label{jordan-wigner}

\begin{figure}
\begin{center}
\resizebox{0.9\columnwidth}{!}{%
  \includegraphics{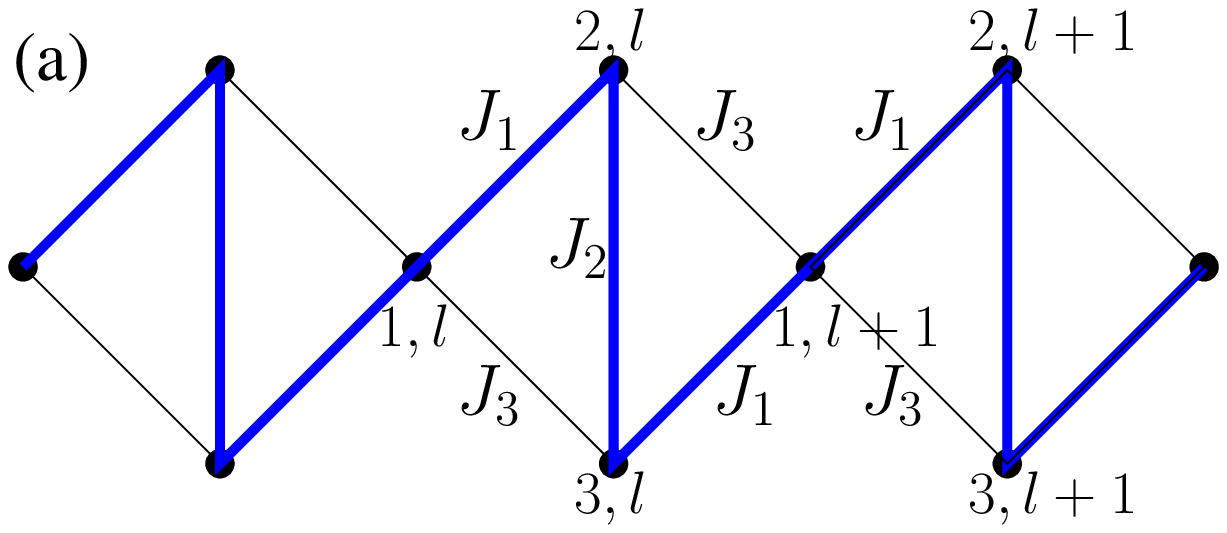}
}\\[12pt]
\resizebox{0.9\columnwidth}{!}{%
  \includegraphics{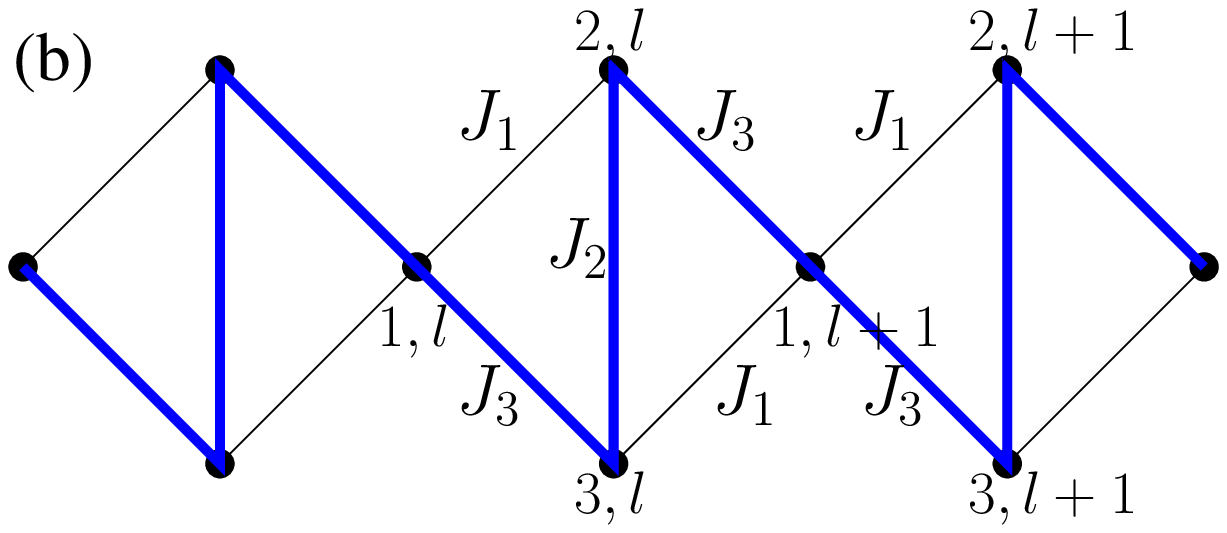}
}
\end{center}
\caption{Different ways of assigning Jordan-Wigner transformation
on the diamond chain are marked by
thick lines. The case {\bf i)} (panel a) and the case {\bf ii)} (panel b), see the main text.}
\label{fig1}
\end{figure}

Consider the quantum spin-$\frac{1}{2}$ $XX$ model on the diamond chain (see Fig.~\ref{fig1})
with the following Hamiltonian:
\begin{eqnarray}
\label{ham-xx}
H_{xx}
&{=}&\frac{1}{2}\sum_{l=1}^N\biggl[
(J_2 s_{2,l}^+ s_{3,l}^-
+ J_1 (s_{1,l}^+ s_{2,l}^-
+  s_{3,l}^+ s_{1,l+1}^-)
\nonumber\\
&&\left.
+ J_3 (s_{1,l}^+ s_{3,l}^-
+  s_{2,l}^+ s_{1,l+1}^-)
+ \mbox{h.c.})
\right.
\nonumber\\
&&
-2h\sum_{p=1}^3\left(s_{p,l}^+s_{p,l}^- -\frac{1}{2}\right) \biggr].
\end{eqnarray}
Here,
$N$ is the number of unit cells,
$s_{p,l}^{\pm}=s_{p,l}^x \pm i s_{p,l}^y$ are the spin raising and lowering operators,
$s_{p,l}^z=s_{p,l}^+s_{p,l}^- - \frac{1}{2}$,
$s_{m,l}^{\alpha}$ ($\alpha=x,y,z$) are the usual Cartesian components
of the Pauli spin-$\frac{1}{2}$ operator with the first index corresponding
to a sublattice and the second to a cell.
$h$ is the external magnetic field (we set $g \mu_B = 1$).
We will further distinguish between two models: the distorted ($J_1\neq J_3$)
and the symmetric ($J_1=J_3$) diamond chain.

The Jordan-Wigner transformation can be unambiguously defined on a linear chain \cite{lsm,katsura},
where all sites can be enumerated subsequently. The diamond chain is however a three sublattice
model and there exists at least two identical ways of arranging its sites in a one-dimensional
sequence as shown on panels (a) and (b) in Fig.~\ref{fig1}. The case {\bf i)} sets the following
order of sites \dots, $(3,l-1)$, $(1,l)$, $(2,l)$, $(3,l)$, $(1,l+1)$, \dots
and the case {\bf ii)} corresponds to another choice of order
\dots, $(2,l-1)$, $(1,l)$, $(3,l)$, $(2,l)$, $(1,l+1)$, \dots.

Following the usual procedure one can define the non-local Jordan-Wigner transformation,
which introduces new Fermi operators by multiplying the spin lowering and raising operators
by the Jordan-Wigner factors $(-2s_{l,m}^z)$
referring to all preceding spins \cite{lsm,katsura}.
For the case {\bf i)} we have:
\begin{eqnarray}
\label{jw1inv}
&&s_{1,l}^-{=}a_{1,l}{\exp}{\left[{-}i\pi
\displaystyle{\sum_{p=1}^3\sum_{i=1}^{l-1}}a_{p,i}^+a_{p,i}\right]},
\nonumber\\
&&s_{2,l}^-{=}a_{2,l}{\exp}{\left[{-}i\pi\left(a_{1,l}^+a_{1,l} {+} \displaystyle{\sum_{p=1}^3\sum_{i=1}^{l-1}}
a_{p,i}^+a_{p,i}
\right)\right]},
\nonumber\\
&&s_{3,l}^-{=}a_{3,l}{\exp}{\left[{-}i\pi\left(a_{1,l}^+a_{1,l} {+} a_{2,l}^+a_{2,l}
{+} \displaystyle{\sum_{p=1}^3\sum_{i=1}^{l-1}}a_{p,i}^+a_{p,i}
\right)\right]},
\nonumber\\
&&s_{n,l}^z=s_{n,l}^+s_{n,l}^- - \frac{1}{2}
=a_{n,l}^+a_{n,l} - \frac{1}{2}.
\end{eqnarray}
Subsequently, the spin interaction terms to emerge in the Hamiltonian (\ref{ham-xx}) can also
be rewritten by the use of Eqs. (\ref{jw1inv}) into the fermionic representation:
$s_{1,l}^+s_{2,l}^-=a_{1,l}^+a_{2,l}$,
$s_{2,l}^+s_{3,l}^-=a_{2,l}^+a_{3,l}$,
$s_{1,l}^+s_{3,l}^-=a_{1,l}^+{\mbox e}^{i\pi a_{2,l}^+a_{2,l}}a_{3,l}$,
\linebreak
$s_{2,l}^+s_{1,l+1}^-=a_{2,l}^+{\mbox e}^{i\pi a_{3,l}^+a_{3,l}}a_{1,l+1}$,
$s_{3,l}^+s_{1,l+1}^-=a_{3,l}^+a_{1,l+1}$.
\linebreak
Hence, it follows that the fermionic representation of the Hamiltonian of
the spin-$\frac{1}{2}$ $XX$ diamond chain then reads:
\begin{eqnarray}
\label{ham-xx-jw1}
H_{xx}=&&\frac{1}{2}\sum_{l=1}^N\biggl[
(J_2 a_{2,l}^+ a_{3,l}
+ J_1 (a_{1,l}^+ a_{2,l} + a_{3,l}^+a_{1,l+1})
\nonumber\\
&&\left.
+ J_3 (a_{1,l}^+ {{\mbox e}^{i\pi a_{2,l}^+a_{2,l}}}a_{3,l}
+ a_{2,l}^+ {{\mbox e}^{i\pi a_{3,l}^+a_{3,l}}}a_{1,l+1})
\right.
\nonumber\\
&&
+\mbox{h.c.})
-2h\sum_{p=1}^3\left(a_{p,l}^+a_{p,l}-\frac{1}{2}\right)
\biggr].
\end{eqnarray}
In the current ordering the sites $(1,l)$, $(3,l)$ and $(2,l)$, $(1,l+1)$
are not nearest neighbors. Therefore, the corresponding couplings contain
phase factors in the hopping terms that prevent a rigorous consideration
of the investigated model.
By imposing periodic boundary conditions in the considered model
$s_{N,l}^{\pm}s_{1,m}^{\mp}$ transform into many fermion terms.
Such boundary terms may be neglected
after making the Jordan-Wigner transformation,
since they are irrelevant for the study of static properties in the thermodynamic limit
\cite{lsm,katsura}.

Considering the numbering of sites {\bf ii)}
the Jordan-Wigner transformation should be defined as:
\begin{eqnarray}
\label{jw2inv}
&&s_{1,l}^-{=}\tilde{a}_{1,l}{\exp}{\left[{-}i\pi
\displaystyle{\sum_{p=1}^3\sum_{i=1}^{l-1}}\tilde{a}_{p,i}^+\tilde{a}_{p,i}\right]},
\nonumber\\
&&s_{2,l}^-{=}\tilde{a}_{2,l}{\exp}{\left[{-}i\pi\left(\tilde{a}_{1,l}^+\tilde{a}_{1,l}
{+}\tilde{a}_{3,l}^+\tilde{a}_{3,l}
{+} \displaystyle{\sum_{p=1}^3\sum_{i=1}^{l-1}}\tilde{a}_{p,i}^+\tilde{a}_{p,i}
\right)\right]},
\nonumber\\
&&s_{3,l}^-{=}\tilde{a}_{3,l}{\exp}{\left[{-}i\pi\left(
\tilde{a}_{1,l}^+\tilde{a}_{1,l}
{+} \displaystyle{\sum_{p=1}^3\sum_{i=1}^{l-1}}\tilde{a}_{p,i}^+\tilde{a}_{p,i}
\right)\right]}.
\end{eqnarray}
Hence, one gets another fermionic representation of the original Hamiltonian
(\ref{ham-xx}):
\begin{eqnarray}
\label{ham-xx-jw2}
H_{xx}=&&\frac{1}{2}\sum_{l=1}^N\bigg[
(J_2 \tilde{a}_{2,l}^+\tilde{a}_{3,l}
+ J_3 (\tilde{a}_{1,l}^+ \tilde{a}_{3,l}
+  \tilde{a}_{2,l}^+ \tilde{a}_{1,l+1})
\nonumber\\
&&\left.
+ J_1 (\tilde{a}_{1,l}^+
{{\mbox e}^{i\pi \tilde{a}_{3,l}^+\tilde{a}_{3,l}}} \tilde{a}_{2,l}
+ \tilde{a}_{3,l}^+
{{\mbox e}^{i\pi \tilde{a}_{2,l}^+\tilde{a}_{2,l}}}\tilde{a}_{1,l+1})
\right.
\nonumber\\
&&
+ \mbox{h.c.})
-2h\sum_{p=1}^3\left(\tilde{a}_{p,l}^+\tilde{a}_{p,l}-\frac{1}{2 }\right)
\bigg].
\end{eqnarray}
It should be noted that the Hamiltonian (\ref{ham-xx-jw2}) differs from
the Hamiltonian (\ref{ham-xx-jw1}). The hopping terms, which have not contained
the fermion interaction before, now acquire a phase factor, and conversely,
in other terms the relevant phase factors have disappeared.
A common feature of both Hamiltonians is that they contain the interaction
only at some particular bonds, and therefore, if one intends to treat the
model approximatively, the contributions of different bonds will not be
generally considered on equal footing.

Using (\ref{jw1inv}), (\ref{jw2inv})
one can check that operators $a_{p,m}$ and $\tilde{a}_{p,m}$ are connected
through the gauge transformation:
\begin{eqnarray}
\label{gauge1}
&&a_{1,l}{=}\tilde{a}_{1,l},
\nonumber\\
&&a_{2,l}{=}\tilde{a}_{2,l}\exp{(i\pi\tilde{a}_{3,l}^+\tilde{a}_{3,l})},
\tilde{a}_{2,l}{=}a_{2,l}\exp{(i\pi a_{3,l}^+a_{3,l})},
\nonumber\\
&&a_{3,l}{=}\tilde{a}_{3,l}\exp{(i\pi\tilde{a}_{2,l}^+\tilde{a}_{2,l})},
\tilde{a}_{3,l}{=}a_{3,l}\exp{(i\pi a_{2,l}^+a_{2,l})}.
\end{eqnarray}
Thus, the Hamiltonians (\ref{ham-xx-jw1}) and (\ref{ham-xx-jw2}) describe
the same physics.

One can also consider a more general gauge transformation
for Fermi operators $c_x$:
\begin{equation}
\label{gauge2}
\tilde{c}_x={\mbox e}^{i\sum_{y\neq x}\alpha_{xy}n_y}c_x,\;
\end{equation}
where $x$, $y$ are general indices which may include the site and sublattice numbers.
If $\alpha_{xy}=\alpha_{yx}+2\pi n$
new operators $\tilde{c}_x$ satisfy the Fermi commutation relations.

Fixing the gauge in transformation (\ref{gauge2}),
we can consider a particular case:
\begin{eqnarray}
\label{gauge3}
&&c_{1,l}=a_{1,l},
\nonumber\\
&& c_{2,l}={\mbox e}^{i\frac{\pi}{2} a_{3,l}^+a_{3,l}}a_{2,l},
a_{2,l}={\mbox e}^{-i\frac{\pi}{2} c_{3,l}^+c_{3,l}}c_{2,l},
\nonumber\\
&&c_{3,l}={\mbox e}^{i\frac{\pi}{2} a_{2,l}^+a_{2,l}}a_{3,l},
a_{3,l}={\mbox e}^{-i\frac{\pi}{2} c_{2,l}^+c_{2,l}}c_{3,l}.
\end{eqnarray}
This gives as a result the Hamiltonian in the most symmetric form:
\begin{eqnarray}
\label{ham-xx-jw4}
H_{xx}=&&\frac{1}{2}\sum_{l=1}^N\Bigg[
(
J_1 (c_{1,l}^+ {{\mbox e}^{-i\frac{\pi}{2} c_{3,l}^+c_{3,l}}} c_{2,l}
+c_{3,l}^+{{\mbox e}^{i\frac{\pi}{2}c_{2,l}^+c_{2,l}}}c_{1,l+1})
\nonumber\\
&&
+J_3 (c_{1,l}^+{{\mbox e}^{i\frac{\pi}{2} c_{2,l}^+c_{2,l}}}c_{3,l}
+ c_{2,l}^+{{\mbox e}^{-i\frac{\pi}{2}c_{3,l}^+c_{3,l}}}c_{1,l+1})
\nonumber\\
&&
+J_2 c_{2,l}^+ c_{3,l}+ \mbox{h.c.})
-2h\sum_{p=1}^3\left(c_{p,l}^+c_{p,l}-\frac{1}{2}\right)
\Bigg].
\end{eqnarray}

One can also easily find the spin-fermion transformation that leads to the representation
(\ref{gauge3}). It takes the form:
\begin{eqnarray}
\label{jw3inv}
&&\!\!\!\!\!\!\!{s_{1,l}^-{=}c_{1,l}{\exp}{\left[-i\pi
\displaystyle{\sum_{p=1}^3\sum_{i=1}^{l-1}}c_{p,i}^+c_{p,i}
\right]},
}
\\
&&\!\!\!\!\!\!\!{s_{2,l}^-{=}c_{2,l}{\exp}{\left[{-}i\pi\left(\frac{1}{2} c_{3,l}^+c_{3,l}
{+}c_{1,l}^+c_{1,l}
{+} \displaystyle{\sum_{p=1}^3\sum_{i=1}^{l-1}}c_{p,i}^+c_{p,i} \right) \right]},
}
\nonumber\\
&&\!\!\!\!\!\!\!{s_{3,l}^-{=}c_{3,l}{\exp}{\left[{-}i\pi\left(-\frac{1}{2}c_{2,l}^+c_{2,l}
{+}c_{1,l}^+c_{1,l}
{+} \displaystyle{\sum_{p=1}^3\sum_{i=1}^{l-1}}c_{p,i}^+c_{p,i}\right)\right]}.
}
\nonumber
\end{eqnarray}
Finally, one should note that equations (\ref{jw3inv})
represent a more general formulation
of the Jordan-Wigner transformation for the spin ladders:
\begin{eqnarray}
\label{jw-gen}
s_{x}^+=c_x^+{\mbox e}^{i\pi\phi_x},
\phi_x=\sum_{y\neq x} \varphi_{x,y}c_y^+ c_y
\end{eqnarray}
with the condition $\varphi_{x,y}=\varphi_{y,x}\pm 1$ \cite{nunner}.
Here $x$, $y$ are again general indices which may include the site and sublattice numbers.

We would like also to comment some limiting cases of the model.
If $J_1=0$ or $J_3=0$ the model resembles
the exactly solvable trimerized $XX$ chain \cite{zaspel1987,okamoto1992,okamoto1996}.
Indeed, if $J_3=0$ (or $J_1=0$) we come after the Jordan-Wigner transformation
(\ref{jw1inv}) (or (\ref{jw2inv})) to the Hamiltonian
(\ref{ham-xx-jw1}) (or (\ref{ham-xx-jw2})) in the form of free fermions.
In case $J_2=0$ but $J_1\neq 0$, $J_3\neq 0$ it is not possible
to find such a form of the Jordan-Wigner transformation
that avoids the interaction between fermions in the transformed Hamiltonian.
Therefore, the consideration of the models with $J_1\neq 0$, $J_3\neq 0$
requires further approximations even if $J_2=0$.

For the further consideration
it is useful to remind that
the $| \downarrow\rangle_n$ ($| \uparrow\rangle_n$) spin state
corresponds to the empty $|0\rangle_n$ (filled $|1\rangle_n=c_n^+|0\rangle_n$) fermion state.
Thus, the magnetization per spin in $z$-direction is related to the fermionic averages
in the following way:
\begin{eqnarray}
\label{magnet}
m_z{=}\frac{1}{3N}\sum_{p=1}^3\sum_{l=1}^N\left\langle s_{p,l}^z\right\rangle
{=}\frac{1}{3N}\sum_{p=1}^3\sum_{l=1}^N\left\langle c_{p,l}^+c_{p,l}\right\rangle{-}\frac{1}{2}.
\end{eqnarray}

\section{Phase factors and their importance. Free-fermion model on diamond chain}
\label{free}

Now we would like to analyze the general properties of the free-fermion model
on the diamond chain, i.e., the model which one gets after the Jordan-Wigner transformation
ignoring the phase factors terms \cite{fu}.

The Hamiltonian of the spin-$\frac{1}{2}$ $XX$ model, as well as a more general $XXZ$ model,
in zero field is invariant under the reflection of $z$-component of all spins:
$s_{p,l}^z\to -s_{p,l}^z$ [$|\uparrow \rangle_{p,l}\to |\downarrow \rangle_{p,l}$].
In fermionic language it corresponds to the particle-hole transformation:
$a_{p,l}\to a_{p,l}^+$ [$|0 \rangle_{p,l}\to |1 \rangle_{p,l}$].
It is clear that the Hamiltonian of the free-fermion gas is not invariant under such transformation
because all the terms change their signs:
\begin{eqnarray}
\label{ham-free-ferm}
H_{ff}=&&-\frac{1}{2}\sum_{l=1}^N\left(
J_2 {a}_{2,l}^+ {a}_{3,l}
+ J_1 ({a}_{1,l}^+ {a}_{2,l}
+ {a}_{3,l}^+ {a}_{1,l+1} )
\right.
\nonumber\\
&&\left.
+ J_3 ({a}_{1,l}^+ {a}_{3,l}
+ {a}_{2,l}^+ {a}_{1,l+1})
+ \mbox{h.c.}
\right).
\end{eqnarray}
One may also use a kind of phase transformation ($\tilde{a}_{p,l}=(-1){a}_{p,l}$)
for some sites to revert the signs.
However, five terms of the Hamiltonian (\ref{ham-free-ferm}) should reverse their signs
for every particular $l$ using transformations for only three operators
and
hence it follows that it is impossible to satisfy simultaneously all the
conditions quite generally.
However, such a possibility appears for some particular cases
$J_1=0$ or $J_3=0$ which correspond to the exactly solvable trimerized chain
\cite{zaspel1987,okamoto1992,okamoto1996}.
The situation for the Hamiltonians (\ref{ham-xx-jw1}) and (\ref{ham-xx-jw2}) is different.
After the particle-hole transformation the terms which contain phase factors preserve their signs,
the signs of other two-fermionic terms can be always reverted
by the aforementioned phase transformations.

Thus, the symmetry of the spin-$\frac{1}{2}$ model is broken in the free-fermion Hamiltonian.
It may mean that the magnetization calculated from this Hamiltonian
is not an odd function of the external magnetic field anymore.
We can prove it in a way similar to \cite{derzhko1999}.
One has to calculate the magnetization using the fermionic representation:
\begin{eqnarray}
m_z{=}\frac{1}{3N}\sum_{p=1}^3\sum_{i=1}^N\langle s_{p,i}^z\rangle
{=}{-}\frac{1}{2}\int\limits_{-\infty}^{\infty}dE\rho(E)\tanh\left(\frac{\beta E}{2}\right)
\label{m_z1}
\end{eqnarray}
where $\beta=1/T$ is the inverse temperature.
We have also introduced the density of states
\begin{eqnarray}
\label{dof}
\rho(E)=\frac{1}{3N}\sum_{p=1}^3\sum_{i=1}^N\delta(E-\Lambda_{p,i}).
\end{eqnarray}
$\Lambda_{p,i}$ denote the eigenvalues of the free-fermion Hamiltonian (\ref{ham-free-ferm}).
Generally, one has to prove the asymmetry of the density of states for $h=0$.
The simplest way is to show that the odd moments for the density of states
\begin{eqnarray}
M^{(l)}=\int_{-\infty}^{\infty} dE E^l \rho(E)
\nonumber
\end{eqnarray}
can take non-zero values.
Using the Green's function method \cite{elk},
we find the third moment in the form:
\begin{eqnarray}
&&M^{(3)}=-h^3 - \frac{1}{2}h(J_2^2{+}2J_1^2{+}2J_3^2) + \frac{1}{2}J_1 J_2 J_3.
\end{eqnarray}
Hence, it remains nonzero in the zero-field case
and gives a nonzero contribution to the zero-field magnetization.

\begin{figure}
\begin{center}
\resizebox{0.9\columnwidth}{!}{%
  \includegraphics{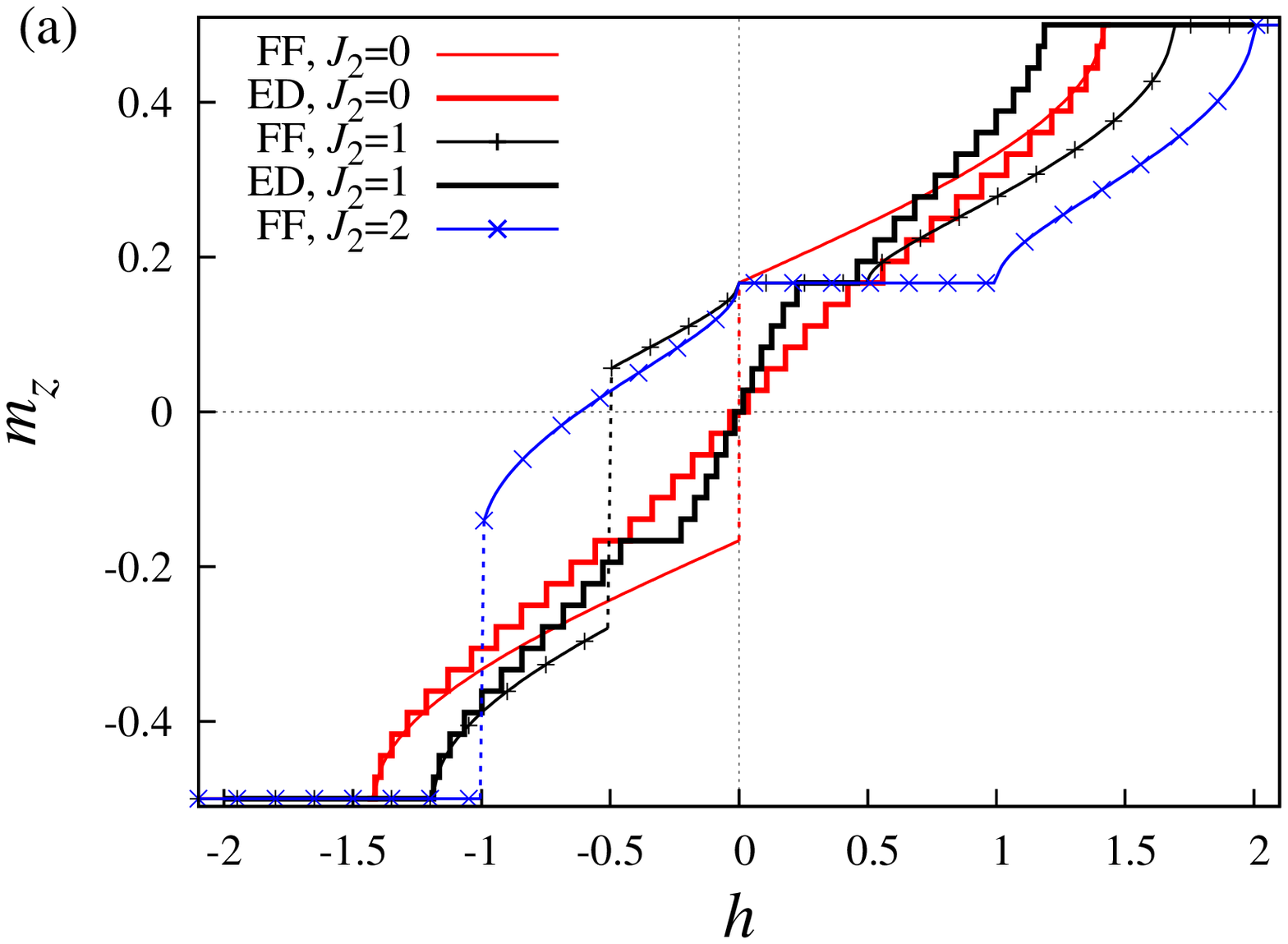}
}
\resizebox{0.9\columnwidth}{!}{%
  \includegraphics{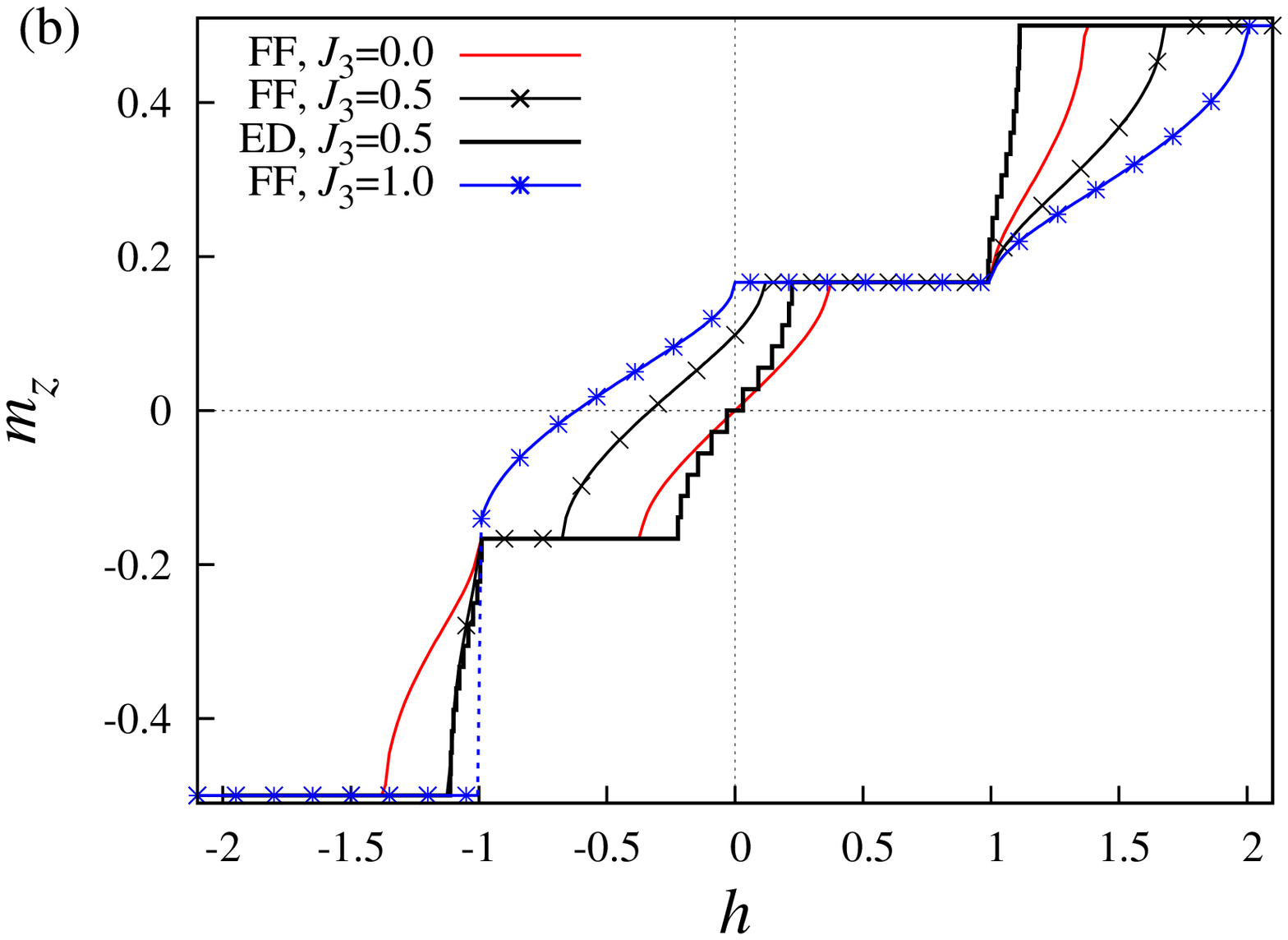}
}
\end{center}
\caption{(Color online) The ground-state magnetization vs. the external field of
the free-fermion (FF) model on:
a) the symmetric diamond chain $J_1=J_3=1$, $J_2=0,1,2$,
the thick step-like lines correspond to the exact diagonalization (ED)
of the spin model (\ref{ham-xx}) of 36 spins for $J_2=0, 1$;
b) the distorted diamond chain $J_2=2$, $J_1=1$, $J_3=0, 0.5, 1$,
the thick step-like line corresponds to the exact diagonalization
of the spin model of 36 spins (\ref{ham-xx}) for $J_3=0.5$.
}
\label{fig_gsmag_ff}
\end{figure}

This general conclusion can be confirmed by numerical calculations.
The free-fermion Hamiltonian (\ref{ham-free-ferm}) can be diagonalized
by subsequent application of the Fourier transformation
(which present the Hamiltonian as a $3\times 3$ quadratic form)
and unitary transformation.
The eigenenergies $\Lambda_n(\kappa)=h+\omega_n(\kappa)$ are expressed through
the solutions $\omega_n(\kappa)$ of the cubic equation:
\begin{eqnarray}
&&\omega_n(\kappa)^3+p(\kappa)\omega_n(\kappa)+q(\kappa)=0,
\nonumber\\
&&p(\kappa)=-\frac{J_2^2}{4}-\frac{J_1^2}{2}-\frac{J_3^2}{2}-J_1 J_3\cos\kappa,
\nonumber\\
&&q(\kappa)=-\frac{J_2}{4}(2J_1 J_3+(J_1^2+J_3^2)\cos\kappa).
\nonumber
\end{eqnarray}
Here $\kappa$ is the wave vector of the Fourier transformation, and $n$ is the sublattice index.
The density of states is rewritten as
$\rho(E)=\sum_n\rho_n(E)$ where $\rho_n(E)=\sum_\kappa \delta(E-h-\omega_n(\kappa))$.
Using this representation of the density of states
and the expression for the magnetization (\ref{m_z1})
one can easily understand the curves shown in Fig.~\ref{fig_gsmag_ff}.
The gap between nearest bands corresponds to the plateau of the magnetization.
The jumps of the magnetization is a mark of the flat band in the model.
Indeed, for the symmetric chain one can find the solution in simple form:
$\omega_1(\kappa)=-J_2/2$,
$\omega_{2,3}(\kappa)=J_2(1\pm\sqrt{1+(4J_1/J_2)^2(1+\cos\kappa)})/4$.
Thus, the ground-state magnetization has the jump at $h=-J_2/2$.
Note that in case all couplings $J_p$ are non-zero, the magnetization gains
a finite value even for zero fields (see Fig.~\ref{fig_gsmag_ff}),
which is clearly an artifact of the
used free-fermion approximation.
In Fig.~\ref{fig_gsmag_ff} we also compare the magnetization
derived from the free-fermion model (\ref{ham-free-ferm})
with the exact diagonalization data of model (\ref{ham-xx}).
One can see that the good agreement is achieved for the strong negative fields only
where the system is close to the ordered state with all spins down.

To conclude, the free-fermion model on the diamond chain generally looses the symmetry
of the corresponding quantum spin model and, moreover, it gives physically not acceptable
solution for the nonzero magnetization even in the zero-field case.

\section{Hartree-Fock approximation for the distorted diamond chain}
\label{distorted_chain_hf}

When the sites ($2,l$) and ($3,l$) are shifted from its symmetric positions,
we come to a distorted diamond chain where only some of the spin interactions
differ from each other.
Without loss of the generality one may impose that $J_1>J_3$ and
choose such a fermionic representation of the Hamiltonian (\ref{ham-xx-jw1})
that the fermion interaction terms
will finally appear along the weaker bonds only. The latter terms can be
then treated perturbatively. Using the algebra of Fermi operators,
i.e., $\exp(i\pi a_{p,l}^+ a_{p,l})=1-2a_{p,l}^+ a_{p,l}$,
one can rewrite (\ref{ham-xx-jw1}) in the following form:
\begin{eqnarray}
\label{ham-xx-jw1a}
H_{xx}&{=}&\frac{1}{2}\sum_{l=1}^N\biggl[
(J_2 a_{2,l}^+ a_{3,l}
+ J_1 (a_{1,l}^+ a_{2,l}
+ a_{3,l}^+a_{1,l+1})
\nonumber\\
&+&\left.
 J_3 (a_{1,l}^+ (1{-} 2a_{2,l}^+a_{2,l})a_{3,l}
{+} a_{2,l}^+ (1{-} 2a_{3,l}^+a_{3,l})a_{1,l+1})
\right.
\nonumber\\
&+&
  \mbox{h.c.})
-2h\sum_{p=1}^3\left(a_{p,l}^+a_{p,l}-\frac{1}{2}\right)
\biggr].
\end{eqnarray}
Applying the Hartree-Fock or mean-field-like
approximation \cite{dmitriev2002,caux2003,hagemans2005}
which preserves the correlations between neighboring sites, e.g.
\begin{eqnarray}
\label{hf}
a_{1,l}^+a_{3,l}^+a_{3,l}a_{2,l}
\approx& -a_{1,l}^+a_{3,l}A_1
-A_3 a_{3,l}^+a_{2,l}
+A_3 A_1
\nonumber\\
&+a_{1,l}^+a_{2,l}\overline{n}_2
+A_2 a_{3,l}^+a_{3,l}
-A_2\overline{n}_2,
\end{eqnarray}
we come to the quadratic form in Fermi operators:
\begin{eqnarray}
\label{ham-xx-jw1b}
H_{xx}&{\approx}&\frac{1}{2}\sum_{l=1}^N\Bigl[ \bigl(
(J_2+4J_3 A_2)a_{2,l}^+ a_{3,l}
\nonumber\\
&&\left.
+ (J_1+2J_3 A_1) (a_{1,l}^+ a_{2,l}
+ a_{3,l}^+a_{1,l+1})
\right.
\nonumber\\
&&\left.
+ J_3(1-2\overline{n}_2) (a_{1,l}^+ a_{3,l}
+ a_{2,l}^+ a_{1,l+1})
+ \mbox{h.c.}\bigr)
\right.
\nonumber\\
&&
-2ha_{1,l}^+a_{1,l}
-2(h+J_3 A_3)(a_{2,l}^+a_{2,l}+a_{3,l}^+a_{3,l})\Bigr]
\nonumber\\
&&
+N e_0.
\end{eqnarray}
Here we use the following notations:
$A_1=\langle a_{2,l}^+a_{3,l}\rangle$,
$A_2=\langle a_{1,l}^+a_{2,l}\rangle=\langle a_{3,l}^+a_{1,l+1}\rangle$,
$A_3=\langle a_{1,l}^+a_{3,l}\rangle=\langle a_{2,l}^+a_{1,l+1}\rangle$,
$\overline{n}_2=\langle a_{2,l}^+a_{2,l}\rangle=\langle a_{3,l}^+a_{3,l}\rangle$,
$e_0=\frac{3h}{2}-4J_3A_1A_2+4J_3\overline{n}_2A_3$.
In the elementary contractions $A_2$, $A_3$ the invariance of
the initial Hamiltonian with respect to the space reflection was exploited.

After Fourier transformation,
\begin{eqnarray}
a_{p,\kappa}=\frac{1}{\sqrt{N}}\sum_{l=1}^N{\mbox e}^{-il\kappa}a_{p,l}, \;
a_{p,l}=\frac{1}{\sqrt{N}}\sum_{\kappa}{\mbox e}^{il\kappa}a_{p,\kappa},
\nonumber\\
\kappa=2\pi m/N, \; m=-\frac{N}{2}+1,\dots, \frac{N}{2} (\mbox{for even}\, N),
\nonumber\\
m=-\frac{N-1}{2},\dots, \frac{N-1}{2} (\mbox{for odd}\, N),
\nonumber
\end{eqnarray}
the Hamiltonian can be represented
in the matrix form
\begin{eqnarray}
&&H_{xx}=\sum_{\kappa}\sum_{p,q=1}^3 {\cal H}_{p,q}(\kappa)a_{p,\kappa}^+a_{q,\kappa}
+N e_0,
\\
&&{\cal H}_{11}(\kappa)=-h,
{\cal H}_{22}(\kappa)={\cal H}_{33}(\kappa)=-h-2A_3J_3,
\nonumber\\
&&{\cal H}_{12}(\kappa){=}{\cal H}_{21}^*(\kappa)
{=}\frac{1}{2}(J_1{+}2J_3A_1 {+}J_3(1{-}2\overline{n}_2){\mbox e}^{-i\kappa}),
\nonumber\\
&&{\cal H}_{13}(\kappa){=}{\cal H}_{31}^*(\kappa)
{=}\frac{1}{2}(J_3(1{-}2\overline{n}_2){+}(J_1{+}2J_3A_1){\mbox e}^{-i\kappa}),
\nonumber\\
&&{\cal H}_{23}(\kappa)={\cal H}_{32}(\kappa)=\frac{1}{2}(J_2+4A_2J_3).
\nonumber
\end{eqnarray}
This Hamiltonian can be reduced to a diagonal form:
\begin{eqnarray}
\label{ham_xx_diag}
H_{xx}=\sum_{p,\kappa}\Lambda_p(\kappa)\eta_{p,\kappa}^+\eta_{p,\kappa}
+N e_0,
\end{eqnarray}
by some unitary transformation
$\eta_{p,\kappa}=\sum_{q=1}^3 u_{p,q}(\kappa)a_{q,\kappa}$
where $u_{p,q}(\kappa)$ is a unitary matrix.
Generally, it corresponds to the $3\times 3$ eigenvalue and eigenvector problems.
Finally, we approximately presented the model as the free-fermion gas,
and its thermodynamic and correlation functions can be calculated straightforwardly.
However, the elementary contractions, which are included into the Hamiltonian parameters,
are unknown and have to be found self-consistently:
\begin{eqnarray}
\label{set_eq}
&&\overline{n}_2{=}\langle a_{2,l}^+a_{2,l}\rangle
{=}\sum_{\kappa}\sum_{p}
|u_{2,p}(\kappa)|^2
\langle \eta_{p,\kappa}^+\eta_{p,\kappa} \rangle,
\nonumber\\
&&A_1{=}\langle a_{2,l}^+a_{3,l}\rangle
{=}\sum_{\kappa}\sum_{p}
u_{2,p}^*(\kappa) u_{3,p}(\kappa)
\langle \eta_{p,\kappa}^+\eta_{p,\kappa} \rangle,
\nonumber\\
&&A_2{=}\langle a_{1,l}^+a_{2,l}\rangle
{=}\sum_{\kappa}\sum_{p}
u_{1,p}^*(\kappa) u_{2,p}(\kappa)
\langle \eta_{p,\kappa}^+\eta_{p,\kappa} \rangle,
\nonumber\\
&&A_3{=}\langle a_{1,l}^+a_{3,l}\rangle
{=}\sum_{\kappa}\sum_{p}
u_{1,p}^*(\kappa) u_{3,p}(\kappa)
\langle \eta_{p,\kappa}^+\eta_{p,\kappa} \rangle.
\end{eqnarray}
Here $\langle\dots\rangle$ means the thermodynamic average with the Hamiltonian
(\ref{ham_xx_diag}),
and for the ideal Fermi gas
$\langle \eta_{p,\kappa}^+\eta_{p,\kappa} \rangle=1/(\exp(\beta\Lambda_p(\kappa))+1)$
is the Fermi function.
Thus, the initial statistical mechanical problem for the spin model is turned into
the solution of the set of equations (\ref{set_eq}).
We have solved this problem numerically and then used the formulae for the free-fermion gas
to calculate the thermodynamics and static properties of the model.
Particularly, the equation (\ref{m_z1}) can be used for calculating the total magnetization.

\begin{figure}
\begin{center}
\resizebox{0.9\columnwidth}{!}{%
  \includegraphics{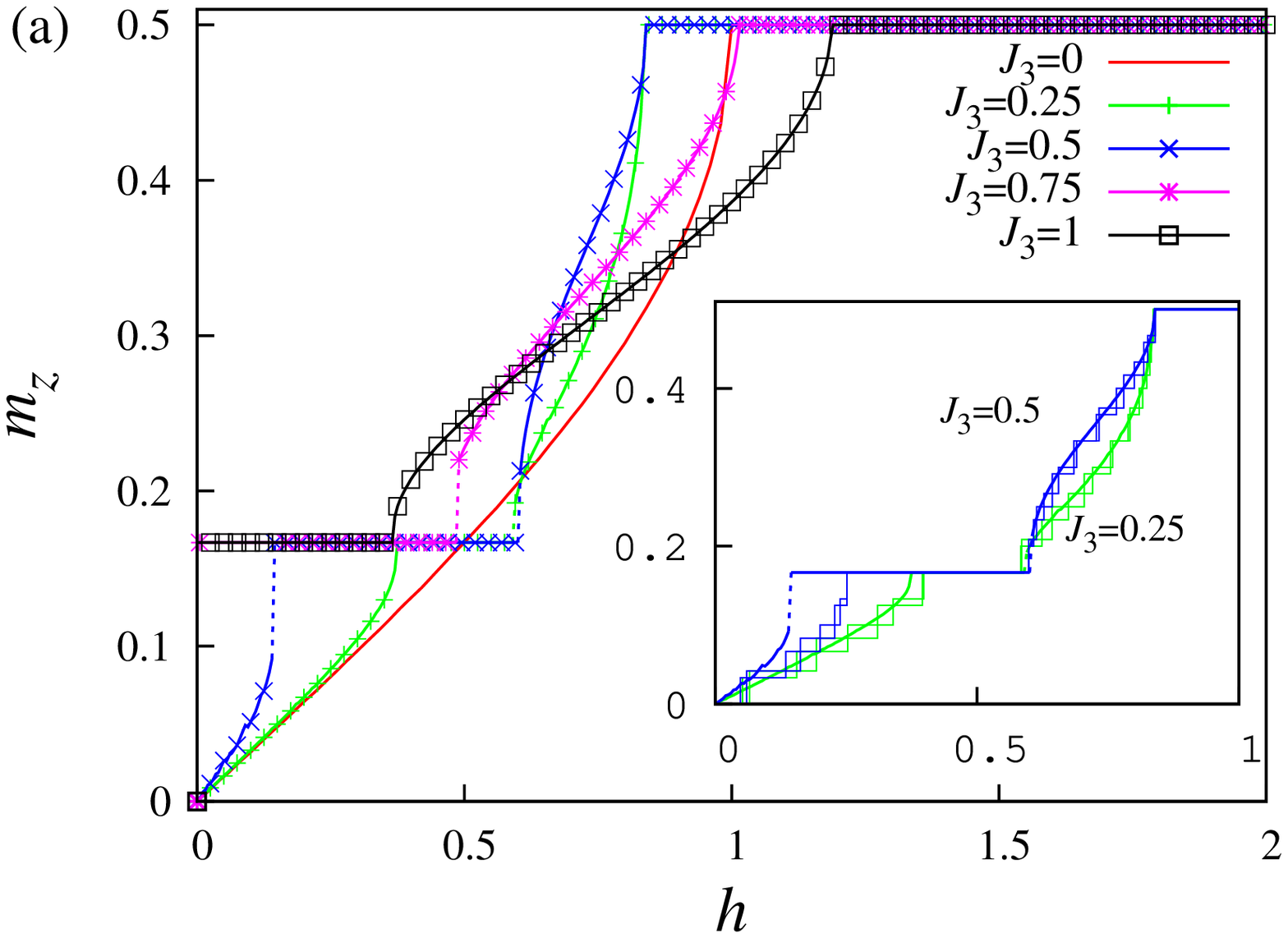}
}
\resizebox{0.9\columnwidth}{!}{%
  \includegraphics{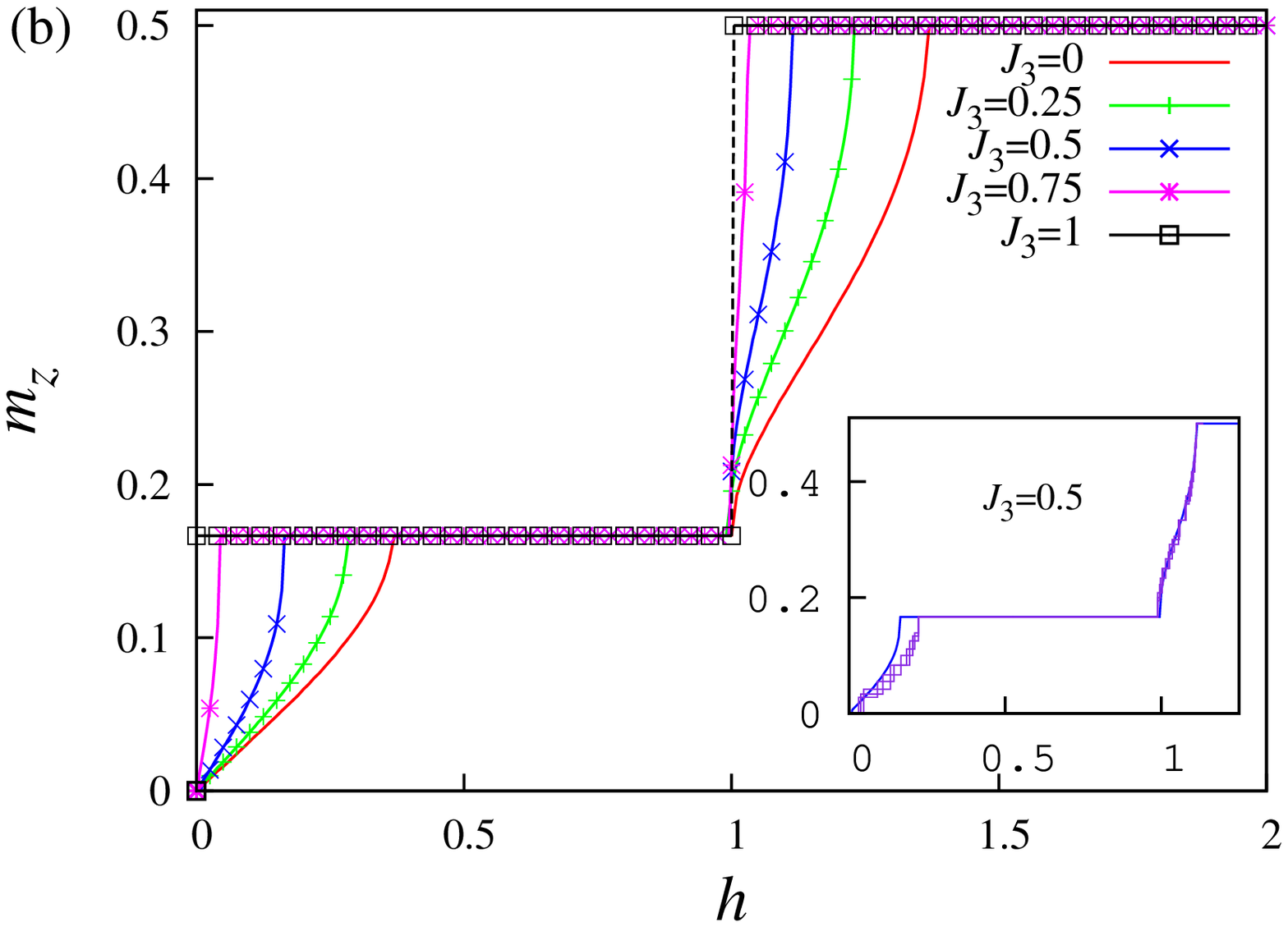}
}
\end{center}
\caption{(Color online) The ground-state magnetization vs. the external field of
the distorted diamond chain for
$J_1=1$, $J_3=0,0.25,0.5,0.75,1$. a) $J_2=1$.
Inset shows the comparison of the results of the Hartree-Fock approximation
and exact diagonalization for $N=24, 30, 36$ (step-like lines) for $J_3=0.25, 0.5$.
b) $J_2=2$.
Inset shows the comparison of the results of the Hartree-Fock approximation
and exact diagonalization for $N=24, 30, 36$ (step-like lines) for $J_3=0.5$.
}
\label{fig_gsmag_dhf}
\end{figure}

For $T=0$ $\langle \eta_{p,\kappa}^+\eta_{p,\kappa} \rangle$ becomes the Heaviside step function
and the summation in (\ref{set_eq}) can be restricted to $\kappa$s which satisfy
$\Lambda_{p}(\kappa)\leq 0$. This simplifies the calculation of the ground state properties of the model.
The results for the ground state magnetization are shown in Fig.~\ref{fig_gsmag_dhf}.
It is worthwhile to remark
that one recovers the exact result for the magnetization curve of the uniform \cite{lsm,katsura}
and trimerized \cite{okamoto1992,okamoto1996}
spin-$\frac{1}{2}$ $XX$ chains in the particular cases
$J_1=J_2=1$, $J_3=0$ and $J_1=1$, $J_2=2$, $J_3=0$
shown in Fig.~\ref{fig_gsmag_dhf}.
Another particular case, which is amenable for an exact calculation, is
the dimer-monomer limit $J_2=2$, $J_1=J_3=1$  (see Fig.~\ref{fig_gsmag_dhf}b).
The magnetization
has a step-like shape in this special case (see Appendix).
The Hartree-Fock equations (\ref{set_eq})
have the following solutions for the elementary contractions:
$\overline{n}_2=\frac{1}{2}$, $A_1=-\frac{1}{2}$, $A_2=A_3=0$ for $h<\frac{J_2}{2}$.
Inserting the solutions obtained for $J_2=2$, $J_1=J_3=1$
into Eq.~(\ref{ham-xx-jw1b}), one comes to
the approximative Hamiltonian in the form
\begin{eqnarray}
\label{ham-hf-dm}
H_{xx}&=&\frac{1}{2}\sum_{l=1}^N \bigg[
J_2(c_{2,l}^+c_{3,l}+c_{3,l}^+c_{2,l})
\nonumber\\
&&-2h\left(c_{1,l}^+c_{1,l}+c_{2,l}^+c_{2,l}+c_{3,l}^+c_{3,l}-\frac{3}{2}\right)\bigg].
\end{eqnarray}
It can be diagonalized by the canonical transformation:
$\eta_{s,l}=\frac{1}{\sqrt{2}}(c_{2,l} {-} c_{3,l})$,
$\eta_{t,l}=\frac{1}{\sqrt{2}}(c_{2,l} {+} c_{3,l})$,
$\eta_{m,l}=c_{1,l}$.
As a result we come to the free-fermion gas of the form:
\begin{eqnarray}
\label{ham-hf-dm2}
H_{xx}&{=}&\frac{1}{2}\sum_{l=1}^N\bigg[
J_2(\eta_{t,l}^+\eta_{t,l} {-}\eta_{s,l}^+\eta_{s,l})
\nonumber\\
&&-2h\bigg(\eta_{t,l}^+\eta_{t,l} {+} \eta_{s,l}^+\eta_{s,l} {+} \eta_{m,l}^+\eta_{m,l} {-} \frac{3}{2}\bigg)\bigg].
\end{eqnarray}
The ground state of the model can be easily found:
$|GS\rangle=\prod_l\frac{1}{\sqrt2}(c_{2,l}^+-c_{3,l}^+)|0\rangle
=\prod_l|\downarrow\rangle_{1,l}[2,l;3,l]$ if $h<0$, \linebreak
$|GS\rangle=\prod_l\frac{1}{\sqrt2}c_{1,l}^+(c_{2,l}^+-c_{3,l}^+)|0\rangle
=\prod_l|\uparrow\rangle_{1,l}[2,l;3,l]$ if $h>0$.
Here $[2,l;3,l]$ represents the singlet dimer state between spins on sites $(2,l)$ and $(3,l)$.
As one can see, the Hamiltonian (\ref{ham-hf-dm2}) indeed represents
the model of non-correlated singlet dimers created between the spins on sites $(2,l)$ and $(3,l)$,
which are separated through
the free monomeric spins residing the sites $(1,l)$ and thus, there cannot be any correlation
between singlet dimers.
We note that the same approach also recovers the completely dimerized state
in the Majumdar-Ghosh limit of the zig-zag ladder \cite{verkholyak2006}.

For the intermediate values of $J_3$ only approximative results are available.
To understand to what extent the elaborated approach is valid
we compare it with the results of the exact diagonalization of finite chain up to 36 spins.
For some particular values of $J_3$
such comparison is shown in the insets to Fig.~\ref{fig_gsmag_dhf}.
It can be seen that the approximative results are quite accurate for sufficiently strong fields $h$,
and one recovers the $1/3$-plateau for the diamond chains.
Therefore, the Hartree-Fock method is reliable even for the case of strong frustration ($J_3\sim J_1$).
However, for small field it may produce an artificial jump of the magnetization
for $J_3=0.25,0.5,0.75$
if one compares the results stemming from the Hartree-Fock approach
with the data of the exact numerical diagonalization method (see. Fig.~\ref{fig_gsmag_dhf}a).

The comparison with the free-fermion model, see Sect.~\ref{free},
demonstrates that neglecting of the phase factors leads to quite
different magnetization curves. Besides the wrong low-field behavior there is
also a significant difference in the saturation field.

The 1/3-plateau in the magnetization curve can be understood from formula (\ref{m_z1})
as an integral over the density of states.
The approximative representation of the initial spin model (\ref{ham-xx})
is a fermionic model with three types of fermions.
Thus, the energy spectrum consists of three bands separated by gaps.
The external field plays the role of the chemical potential of fermions
and thus, it can shift their Fermi level to the area of a gap between two bands.
Such a situation corresponds to the $1/3$-plateau of the magnetization.
It can be also understood that all elementary contractions do not change
along the plateau.
Therefore, the width of the plateau is equal to the width of the gap
in case the magnetization curve does not reveal any jumps.

\begin{figure}[h]
\begin{center}
\resizebox{0.9\columnwidth}{!}{%
  \includegraphics{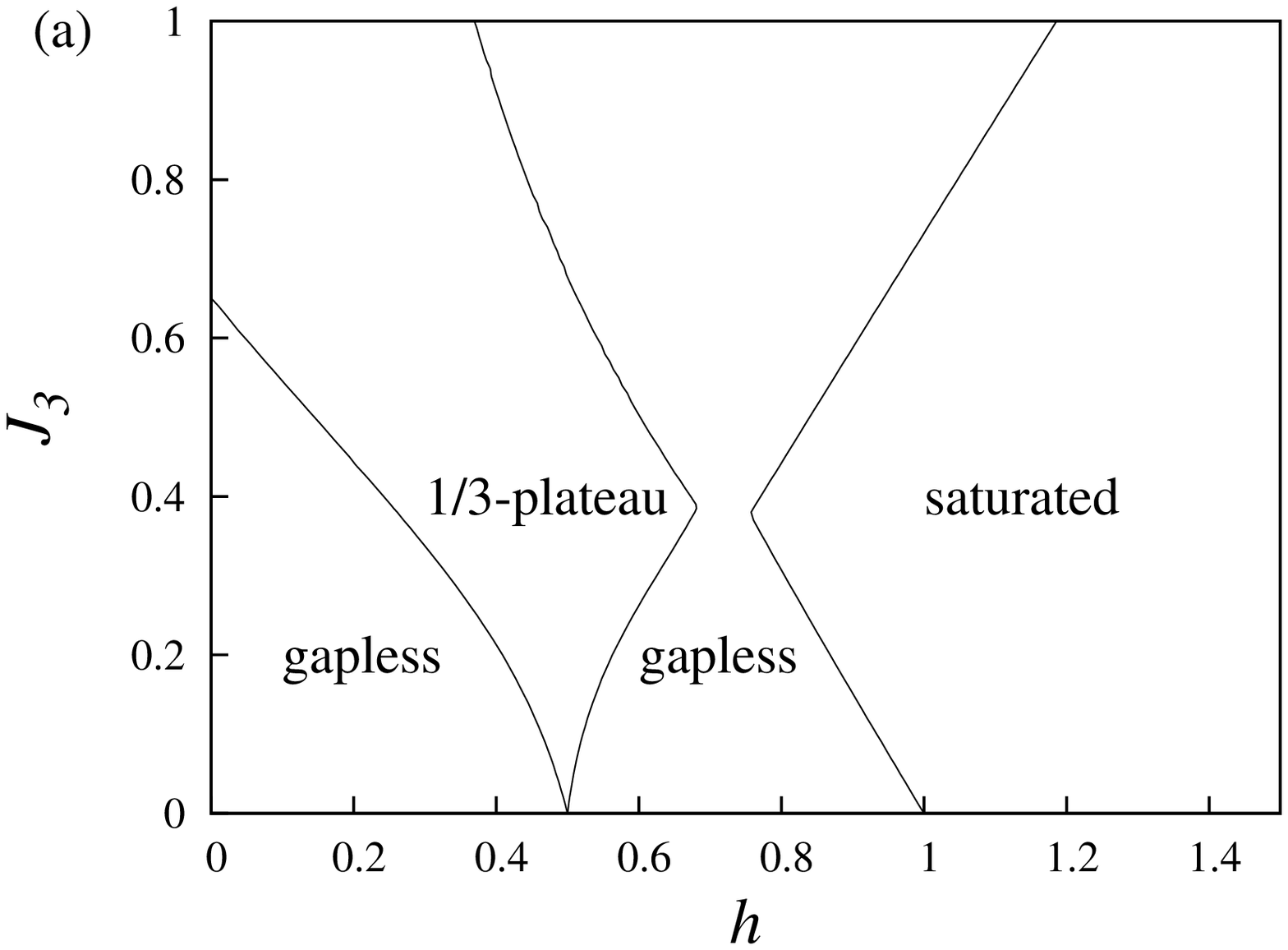}
}
\resizebox{0.9\columnwidth}{!}{%
  \includegraphics{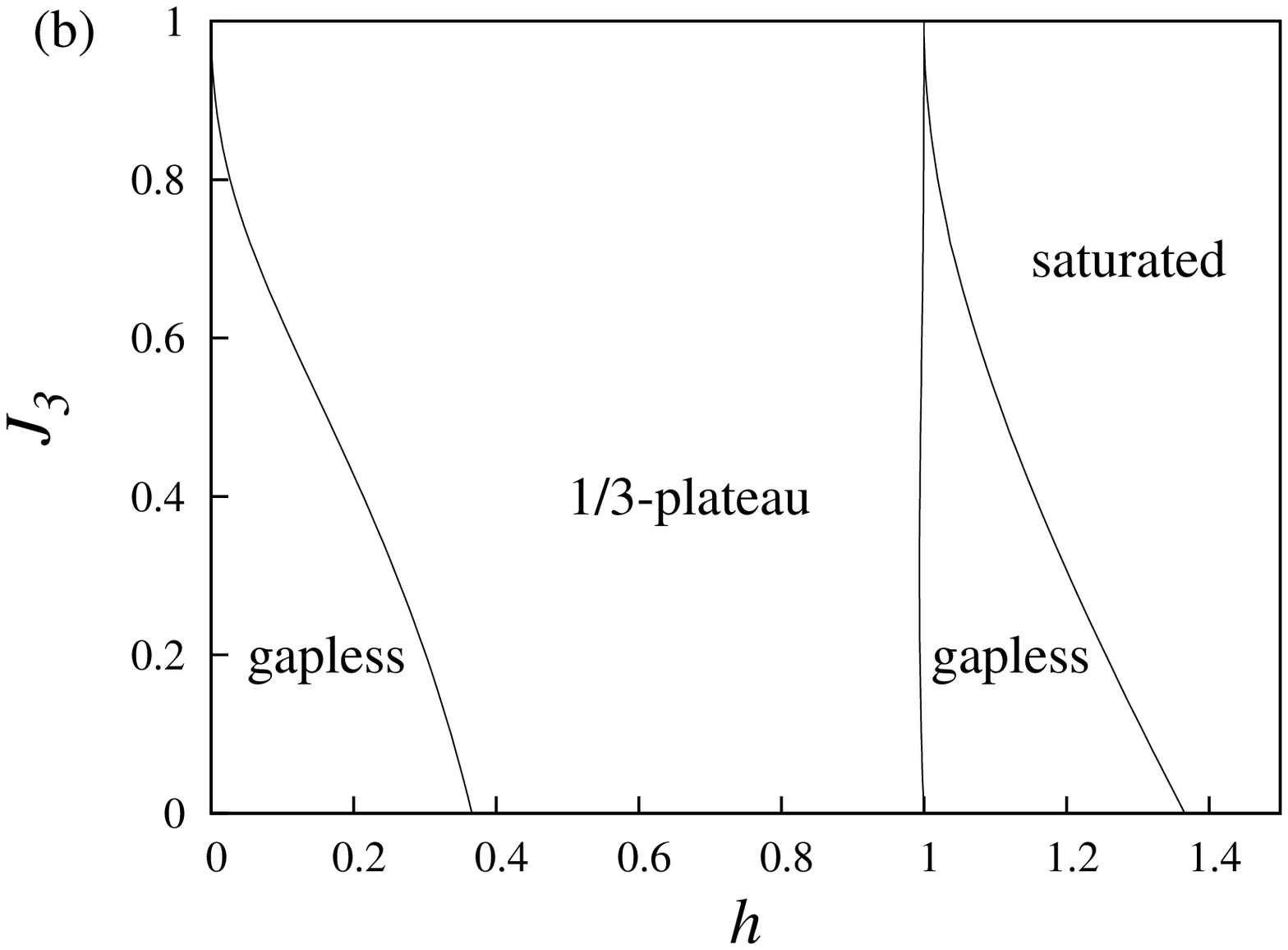}
}
\end{center}
\caption{The ground-state phase diagram of the distorted diamond chain
in the Hartree-Fock approximation.
The upper (lower) panel corresponds to the case $J_1=1$, $J_2=1$ ($J_1=1$, $J_2=2$).}
\label{fig_pd_dist}
\end{figure}

The effect of $J_3$ interaction on the ground-state magnetization process
can be understood from Fig.~\ref{fig_pd_dist}.
We see that for initially uniform chain ($J_2=J_1=1$)
an addition of the frustrating interaction $J_3$ generally causes an onset of the
intermediate magnetization plateau at $m=1/3$, which becomes the broader the stronger
the interaction parameter $J_3$ is (see Fig.~\ref{fig_pd_dist}a).
Generally, the coupling $J_3$ increases the width of the magnetization plateau
also in case of the initial trimerized chain.
It is interesting to note that the upper critical field
for the $1/3$-magnetization plateau in case $J_1=1$, $J_2=2$
does not show any marked dependence on $J_3$.

It should be noted that the ground state phase diagram of
the generally anisotropic distorted diamond chain was studied previously \cite{okamoto2005,okamoto2005-2,okamoto2007}.
The so-called inversion areas may be found in this more general model
where the N\'{e}el phase becomes the ground state in the parameter space with
the predominant $XY$-like interaction and the spin-fluid phase contrary
appears in the parameter region with the predominant Ising-like
interaction.
The  N\'{e}el order considered in Refs.\cite{okamoto2005,okamoto2005-2,okamoto2007}
is related to the antiferromagnetic ordering of spin blocks
$({\bf s}_{3,l-1}+{\bf s}_{1,l}+{\bf s}_{2,l})$,
and thus corresponds to 6 spins in the magnetic cell.
This phase seems to disappear in the pure $XY$ limit
(see Fig.~2 of Ref.\cite{okamoto2007}),
at least it is true for $J_1\gg J_2, J_3$.
Since we have not considered
the possibility of doubling of the elementary magnetic cell in our approach,
we cannot verify whether such a phase exists. Moreover,
the spin-fluid phase found in Ref.~\cite{okamoto2007}
for $h=0$ we do not find in our approach, which might be attributed
to its mean-field character.

\section{Hartree-Fock approximation for the symmetric diamond chain}
\label{symmetric_chain_hf}

Here we consider the case of symmetric diamond chain when the interactions
$J_1=J_3$. For this purpose, let us consider the fermionic representation (\ref{ham-xx-jw4})
of the spin Hamiltonian which provides a symmetric representation for those bonds.
In the fermionic Hamiltonian we keep all the terms that correspond to the correlation
between neighboring sites in the decoupling.
Since the coefficients of the Hamiltonian (\ref{ham-xx-jw4}) are complex
the elementary contractions can also be complex valued.
Similarly to previous section we introduce the following notations:
\begin{eqnarray}
\label{contractions}
&&A_{1}=\langle c_{2,l}^+c_{3,l}\rangle, \;
\nonumber\\
&&A_{2}=\langle c_{1,l}^+c_{2,l}\rangle=\langle c_{1,l+1}^+c_{3,l}\rangle,
\nonumber\\
&&A_{3}=\langle c_{1,l}^+c_{3,l}\rangle=\langle c_{1,l+1}^+c_{2,l}\rangle.
\end{eqnarray}
Here we additionally used the invariance of the initial spin Hamiltonian with the respect
to the space reflection. The invariance to the exchange of the second and third sublattices,
gives us the following connections between different contractions:
$A_2=A_3^*$ ($A_{2}^R=A_{3}^R$, $A_{2}^I=-A_{3}^I$).
Here $A_{p}^R$ ($A_{p}^I$) is the real (imaginary) part of
$A_{p}$.
The Hamiltonian within the adopted approximation will be as follows:
\begin{eqnarray}
\label{ham-hf-sym}
H_{xx}&=&\frac{1}{2}\sum_{l=1}^N [J_2+4J_1(A_2^R+A_2^I)]
(c_{2,l}^+c_{3,l}+c_{3,l}^+c_{2,l})
\nonumber\\
&&+\tilde{J}_1^R
\left(
(c_{1,l}^+ + c_{1,l+1}^+)(c_{2,l}+c_{3,l})+ \mbox{h.c.}
\right)
\nonumber\\
&&+i\tilde{J}_1^I
\left(
(c_{1,l+1}^+ - c_{1,l}^+)(c_{2,l}-c_{3,l})- \mbox{h.c.}
\right)
\nonumber\\
&&
-2(hc_{1,l}^+c_{1,l}{+}h_2c_{2,l}^+c_{2,l}{+}h_2c_{3,l}^+c_{3,l})]
+Ne_0
\end{eqnarray}
where $\tilde{J}_1^R=J_1(1-\overline{n}_2+A_1^R)$,
$\tilde{J}_1^I=J_1(\overline{n}_2+A_1^R)$,
$h_2=h-2J_1(A_2^R-A_2^I)$,
$e_0=\frac{3h}{2}-4J_1A_1^R(A_2^R+A_2^I)+4J_1n_2(A_2^R-A_2^I)$.

After performing Fourier transformation, the Hamiltonian can be written in
a matrix form as follows
\begin{eqnarray}
\label{ham-xx-hf2a}
&&H_{xx}=\sum_{\kappa}\sum_{p,q=1}^3 {\cal H}_{p,q}(\kappa)a_{p,\kappa}^+a_{q,\kappa}
+N e_0
\\
&&{\cal H}_{11}(\kappa)=-h,
\nonumber\\
&&{\cal H}_{22}(\kappa)={\cal H}_{33}(\kappa)=-h-2J_1(A_2^R-A_2^I),
\nonumber\\
&&{\cal H}_{12}(\kappa) {=} {\cal H}_{21}^*(\kappa)
{=} \frac{1}{2}\sqrt{(\tilde{J}_1^R)^2{+}(\tilde{J}_1^I)^2}
({\mbox e}^{-i\phi}{+}{\mbox e}^{i(\phi-\kappa)}),
\nonumber\\
&&{\cal H}_{13}(\kappa) {=} {\cal H}_{31}^*(\kappa)
{=}\frac{1}{2}\sqrt{(\tilde{J}_1^R)^2{+}(\tilde{J}_1^I)^2}
({\mbox e}^{i\phi}{+}{\mbox e}^{-i(\phi+\kappa)}),
\nonumber\\
&&{\cal H}_{23}(\kappa)={\cal H}_{32}(\kappa)=\frac{1}{2}(J_2+4J_1(A_2^R+A_2^I)),
\label{t2t3_hf}
\end{eqnarray}
where $\tilde{J}_1=\sqrt{(\tilde{J}_1^R)^2+(\tilde{J}_1^I)^2}$,
$\tan\phi=\tilde{J}_1^I/\tilde{J}_1^R$.

Similarly to the previous section the problem is rewritten  as the free-fermion gas.
The Hamiltonian (\ref{ham-hf-sym}) or (\ref{ham-xx-hf2a}) contains
the unknown contractions $\overline{n}_2$, $A_1$, $A_2$.
These contractions have to be found from the set of self-consistent equations
identical to (\ref{set_eq}).

\begin{figure}[t]
\begin{center}
\resizebox{0.9\columnwidth}{!}{%
  \includegraphics{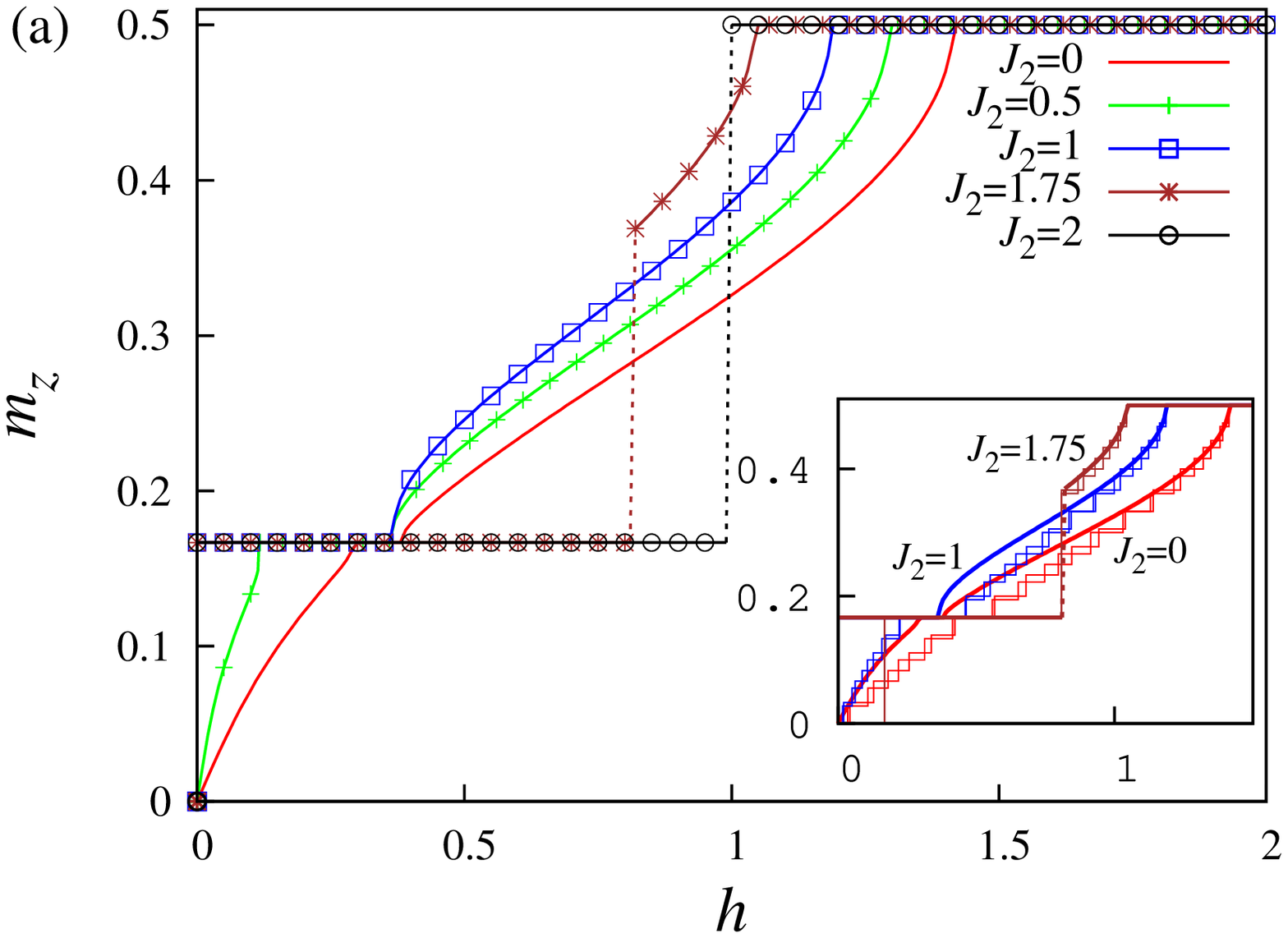}
}
\resizebox{0.9\columnwidth}{!}{%
  \includegraphics{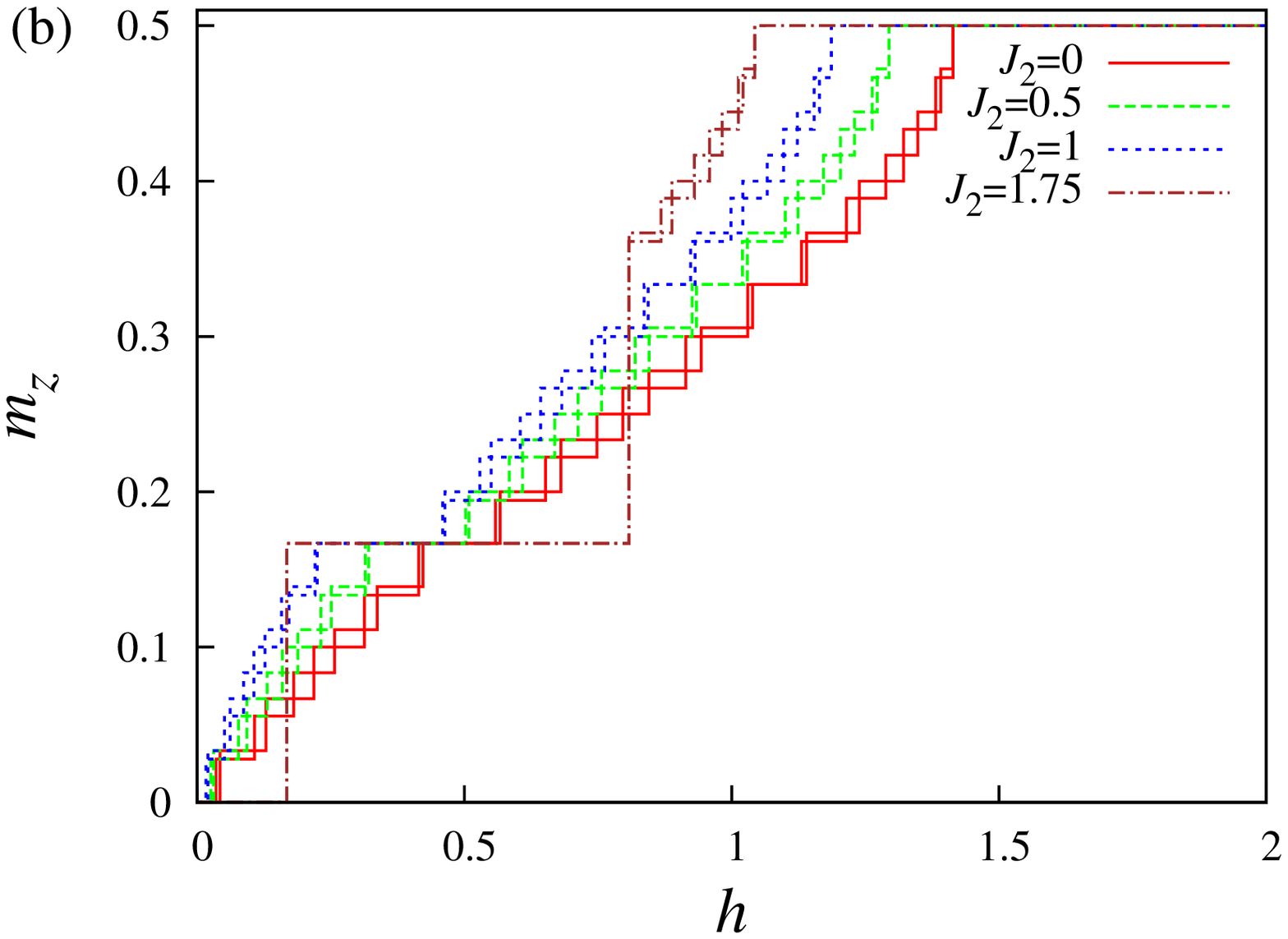}
}
\end{center}
\caption{(Color online) The ground-state magnetization vs. the external field of
the symmetric diamond chain: $J_1=1$
a) the results for the Hartree-Fock approximation are shown
and compared with the exact diagonalization data in the inset for $J_2=0,1,1.75$;
b) the results of the exact diagonalization for finite chains of $N=30,36$ spins
with $J_2=0,0.5,1,1.75$.}
\label{fig_gsmag_symm}
\end{figure}

\begin{figure}[t]
\begin{center}
\resizebox{0.9\columnwidth}{!}{%
  \includegraphics{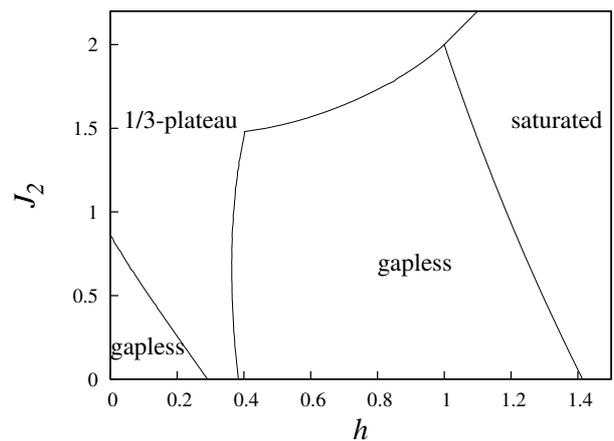}
}
\end{center}
\caption{The ground-state phase diagram in the $h-J_2$ plane
for the symmetric diamond chain with $J_1 = J_3 =1$
in the Hartree-Fock approximation.}
\label{fig_pd}
\end{figure}

The results for the ground-state magnetization are presented in Fig.~\ref{fig_gsmag_symm}.
We set $J_1=1$ here.
One can notice that the magnetization for $J_2=2J_1$ corresponds
to the exact one (see Appendix).
Following the arguments of the previous section,
it can be shown that the Hartree-Fock approximation leads to
the dimer-monomer ground state which is exact for this case.
For $J_2\geq 2J_1$ the magnetization curve still has a step-like shape and corresponds to the exact result.
For intermediate values of $J_2$ we also obtain rather good coincidence
with the exact diagonalization data.
We even recover the magnetization jump for intermediate fields
which is present also in the exact diagonalization data.
However the approximate results for the low-field magnetization
show a noticeable difference with the exact diagonalization data.
Moreover for $J_2\gtrsim 0.86J_1$ we get an incorrect non-zero magnetization even for zero fields $h$.
It is well known that if $0.909<J_2/J_1<2$ the singlet dimer-tetramer phase is the ground state
of the isotropic diamond chain \cite{tks}. This phase survives also in the model
with the easy-plane anisotropy \cite{okamoto2005}.
The elementary magnetic cell is doubled in this case.
Therefore, the current approach, where a doubling of the
unit cell is not taken into account, does not capture
this phase.
At higher fields the dimer-tetramer order is destroyed
and the ordering of the singlet dimers on vertical bonds becomes
more favorable (see Fig.~\ref{fig_gsmag_symm}b).
Since the elementary magnetic cell is not doubled anymore, we obtain an excellent agreement
between the approximate and exact diagonalization data
(see the inset in Fig.~\ref{fig_gsmag_symm}a).
Our exact diagonalization data are also in agreement
with the previous results for the easy-plane $XXZ$ diamond chain \cite{okamoto2005,okamoto2005-2,okamoto2007}.
Note that the model with the easy-plane anisotropy in zero field is characterized
by the spin-fluid phase for $J_2\lesssim J_1$  (see Fig.~3 in Ref.\cite{okamoto2005})
in contrast to the isotropic model where the ferrimagnetic order persists in the ground state \cite{tks}.
The exact diagonalization data show the continuous growth
of the ground-state magnetization with field
for $J_2\lesssim J_1$. This is usually the sign of a gapless excitation spectrum
and the spin-fluid phase.
For stronger $J_2$ couplings we obtain the zero-magnetization plateau and a very steep increase
of the magnetization to the $1/3$ plateau for some critical field ($h\approx0.168J_1$ for $J_2=1.75J_1$).
The phase with the zero magnetization plateau can be identified as the singlet dimer-tetramer phase,
and critical field as a field which destroys tetramers in the ground state.
Considering the ground-state phase diagram presented in Fig.~\ref{fig_pd},
we have to conclude that the considered Hartree-Fock approach is not able
to exhibit the singlet dimer-tetramer phase with the zero magnetization plateau
which appears in the thin region of small fields and $1\lesssim J_2/J_1\leq 2$
in accordance with the exact diagonalization data.
However, as soon as the magnetic cell is not doubled anymore,
the ground state phase diagram are in good agreement with the exact diagonalization data.

Again, the comparison with the free-fermion model, see Sect.~\ref{free},
shows that it is not a reasonable approximation to neglect the phase factors.
In particular, the jump found within our approach and confirmed by exact diagonalization
is not present in the free-fermion model.
Comparing the results of the previous section
we see that both formulations with different gauges produces
the same results for the ground state properties.
Thus the Hartree-Fock approximation preserves the gauge invariance.

The results of the exact diagonalization show also some indication of the second plateau
at $2/3$ of the saturated magnetization.
We note that the elaborated
Hartree-Fock approach cannot reproduce such a behavior.
To get it one has to consider the possibility of
doubling of the elementary magnetic cell in the decoupling procedure
which was not assumed in Eqs. (\ref{contractions}) and (\ref{ham-hf-sym}).

\section{Diamond chain at non-zero temperatures}
\label{temperature_effect}
In this section we consider the thermodynamic properties of the diamond chain
within the fermionization approach and the Hartree-Fock approximation.
After having rewritten the spin Hamiltonian as a quadratic form of Fermi operators,
one can find its eigenvalues and obtain the Helmholtz free energy per site:
\begin{eqnarray}
\label{free-en}
f&=&-\frac{1}{3N\beta}\sum_{p=1}^3\sum_{\kappa}\ln(\mbox{e}^{-\beta\Lambda_p(\kappa)}+1)
\nonumber\\
&=&-\frac{1}{\beta}\int_{-\infty}^{\infty}dE\rho(E)\ln(\mbox{e}^{-\beta E}+1),
\end{eqnarray}
the internal energy per site
\begin{eqnarray}
u=\frac{1}{3N}\langle H\rangle
=\int_{-\infty}^{\infty}dE\rho(E) n(E)E+e_0,
\end{eqnarray}
and the specific heat per site  $c=\frac{du}{dT}$
\begin{eqnarray}
c
{=}\frac{1}{T^2}\int\limits_{-\infty}^{\infty}dE\rho(E) n^2(E)
E^2\mbox{e}^{\beta E}
{+}\frac{\partial u}{\partial \overline{n_2}}
\frac{\partial \overline{n_2}}{\partial T}
{+}\sum_{p=1}^3
\frac{\partial u}{\partial A_p}
\frac{\partial A_p}{\partial T}.
\nonumber\\
\end{eqnarray}
Here $\beta=1/T$ is the inverse temperature.

\begin{figure}[t]
\begin{center}
\resizebox{0.9\columnwidth}{!}{%
  \includegraphics{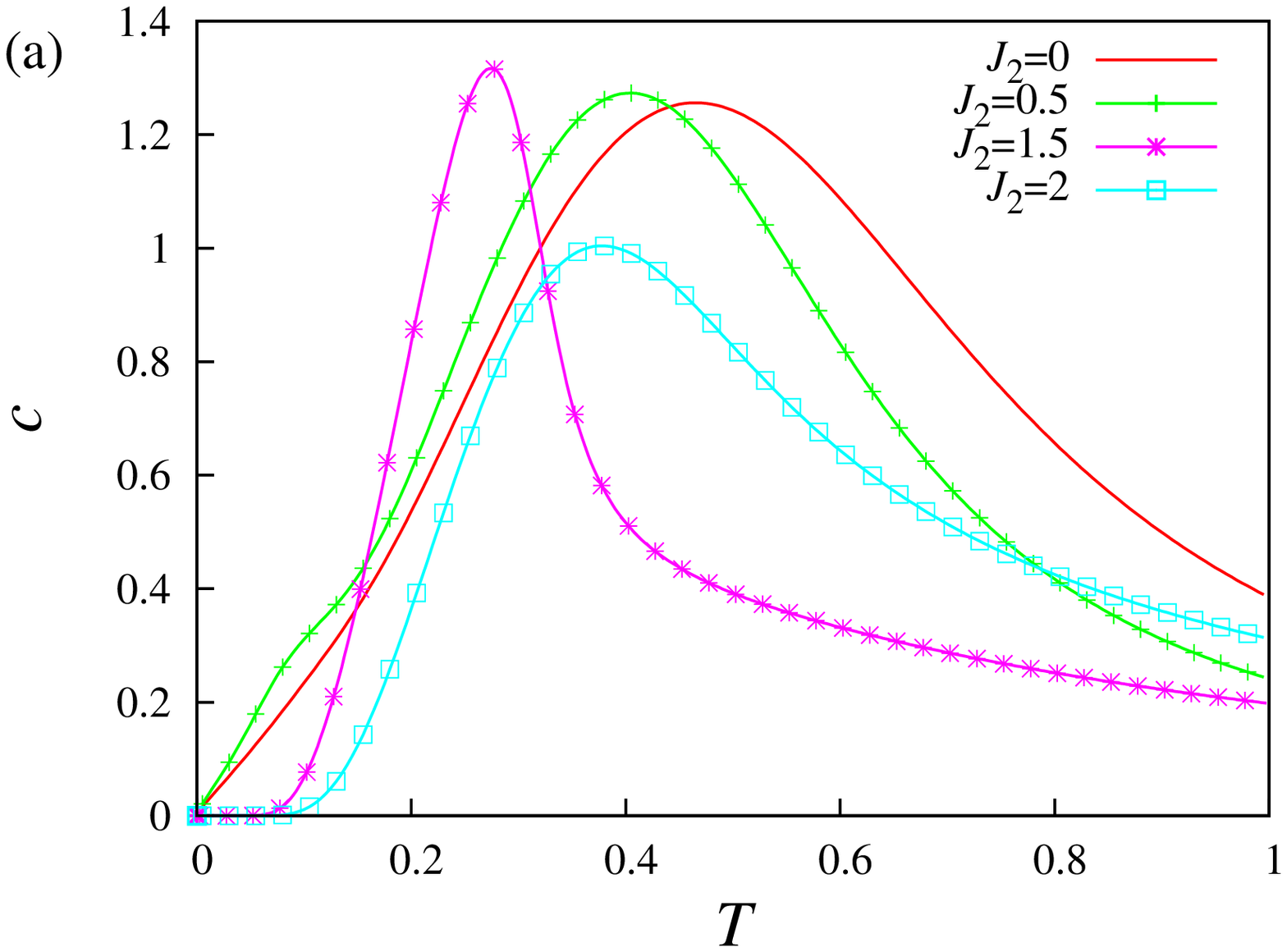}
}
\resizebox{0.9\columnwidth}{!}{%
  \includegraphics{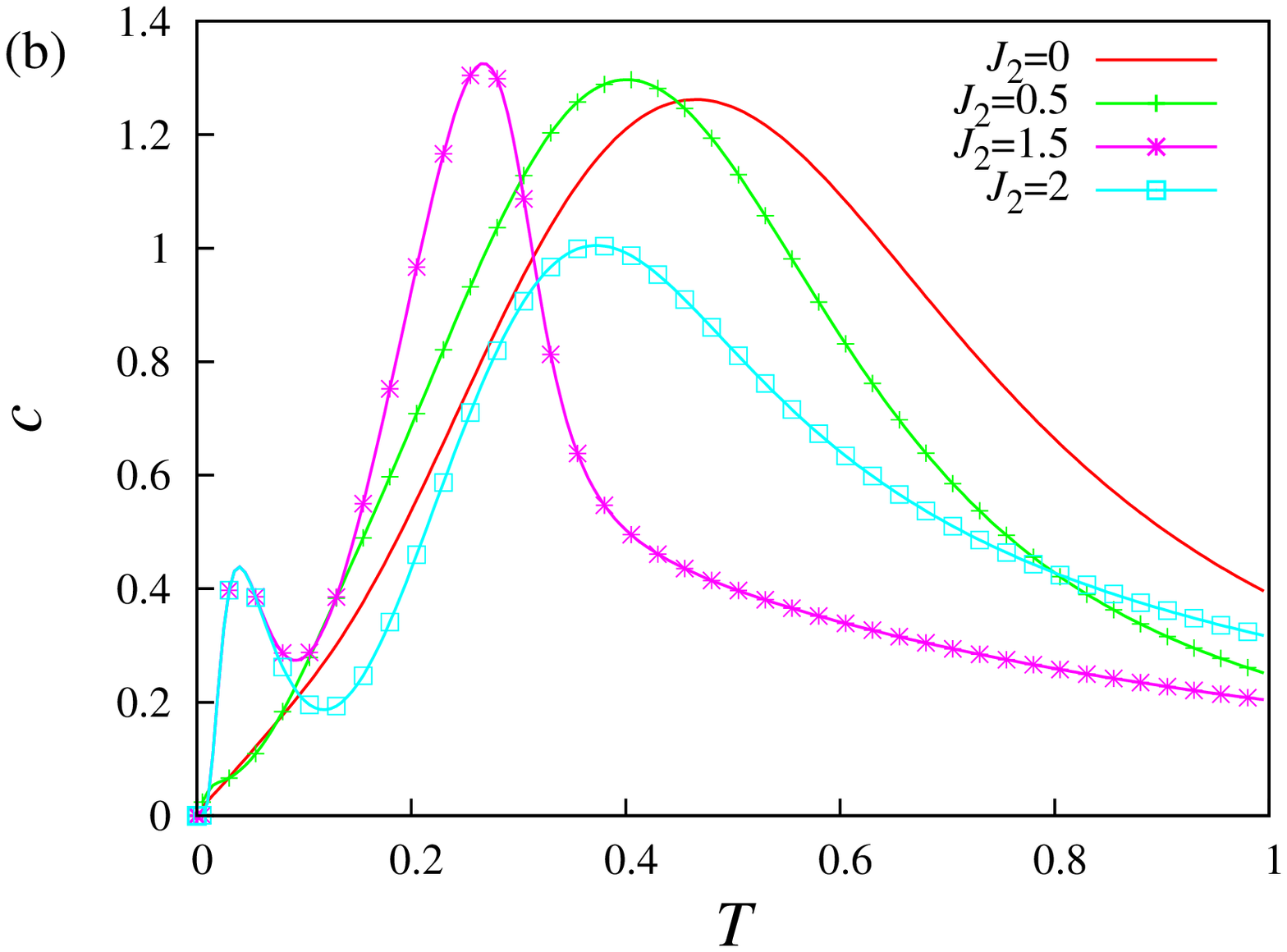}
}
\end{center}
\caption{(Color online) The specific heat in the Hartree-Fock approximation
vs. temperature of
the symmetric diamond chain ($J_1=J_3=1$), $J_2=0, 0.5, 1.5,2$:
a) $h=0$, b) $h=0.1$.}
\label{fig_heat}
\end{figure}

\begin{figure}[t]
\begin{center}
\resizebox{0.9\columnwidth}{!}{%
  \includegraphics{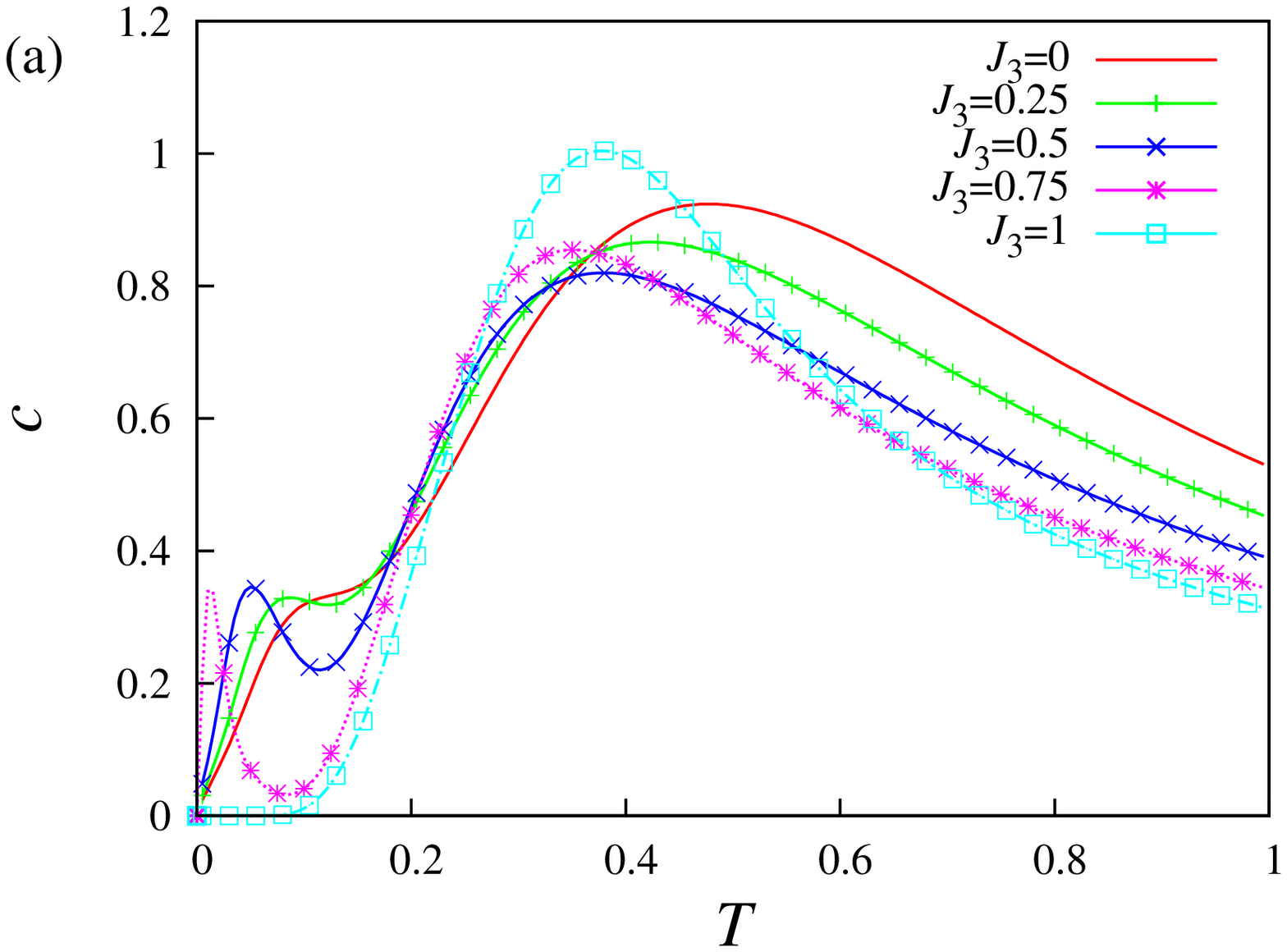}
}
\resizebox{0.9\columnwidth}{!}{%
  \includegraphics{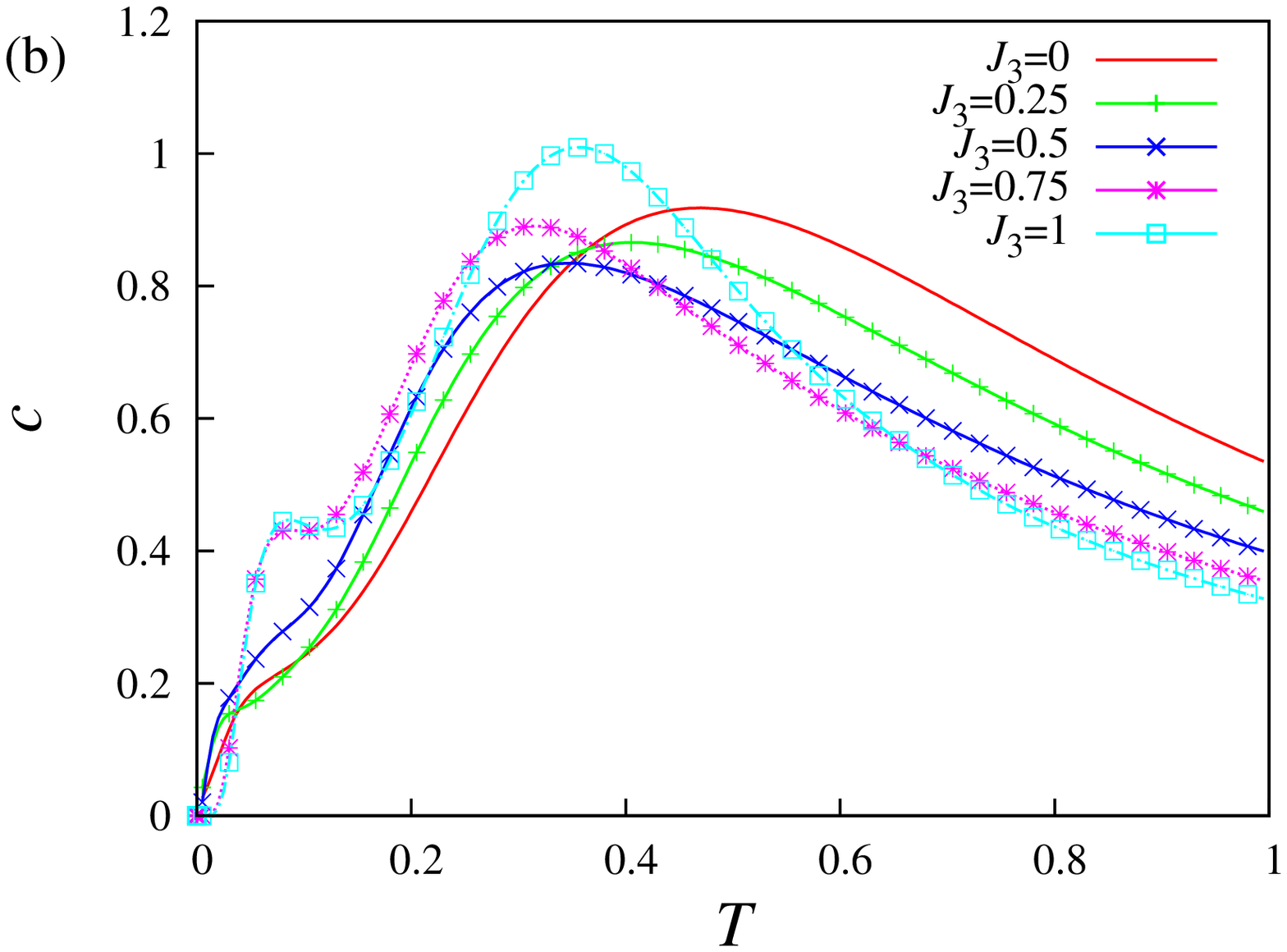}
}
\end{center}
\caption{(Color online) The specific heat in the Hartree-Fock approximation
vs. temperature of
the distorted diamond chain
$J_1=1$, $J_2=2$, $J_3=0, 0.25, 0.5, 0.75, 1$:
a) $h=0$, b) $h=0.2$.}
\label{fig_dheat}
\end{figure}

To calculate the specific heat we perform the numerical differentiation.
The results of computations are shown in Figs.~'¨\ref{fig_heat} and \ref{fig_dheat}.
Some typical features can be observed there.
The specific heat for small vertical couplings $J_2$ grows up linearly with the temperature
which is inherent for systems with a gapless excitation spectrum.
For sufficiently strong $J_2$ (see the curves corresponding to $J_2=1.5,2$  in Fig.~\ref{fig_heat})
the ground state is degenerate due to dimer-monomer phase,
and the system has a gap between the ground and excited states.
Therefore, we observe
the exponential growth of the specific heat with temperature.
When a small external field $h$ is applied, it does not produce a gap in the excitation spectrum,
thus the temperature dependence of the specific heat remains qualitatively the same.
The effect of the external field on the dimer-monomer phase is somewhat different.
The ground state degeneracy is lifted due to the Zeeman term $-h\sum_{p,l}s_{p,l}^z$.
As a results we have two kinds of excitations with different gaps.
This leads to the appearance of the double-peak structure in the
temperature dependence of the specific heat.
The first peak corresponds to the low-energy flip excitations of the monomer spins $s_{1,l}$,
and the second one corresponds
to the thermal excitations of the spins creating singlet dimers.
Here we should emphasize that the dimer-monomer phase as well as the ground state degeneracy
is an artifact of the elaborated approximation for $0.909<J_2/J_1<2$ (see discussion in Sec.~5).
Therefore, the results presented in Fig.~\ref{fig_heat}
for $J_2=1.5$ cannot reproduce
the features of the specific heat of the spin-$\frac{1}{2}$ $XX$ diamond chain.

In contrast to the symmetric chain, the
distorted diamond chain shows the double-peak structure in the temperature
dependence of the specific heat even in the zero-field case (see Fig.~\ref{fig_dheat}),
whereas the external field can even demolish this double-peak structure.
The mentioned features are not only the consequence
of the $XX$ anisotropy in the spin interaction.
The similar behavior
of the temperature dependent specific heat with the double-peak structure
was also reported
for the Heisenberg diamond chain,
see Fig.~8 of Ref.~ \cite{gu}.
The external field can also lead to less visible double-peak structure
in the specific heat of the Heisenberg diamond chain \cite{li2007}.

\begin{figure}[t]
\begin{center}
\resizebox{0.9\columnwidth}{!}{%
  \includegraphics{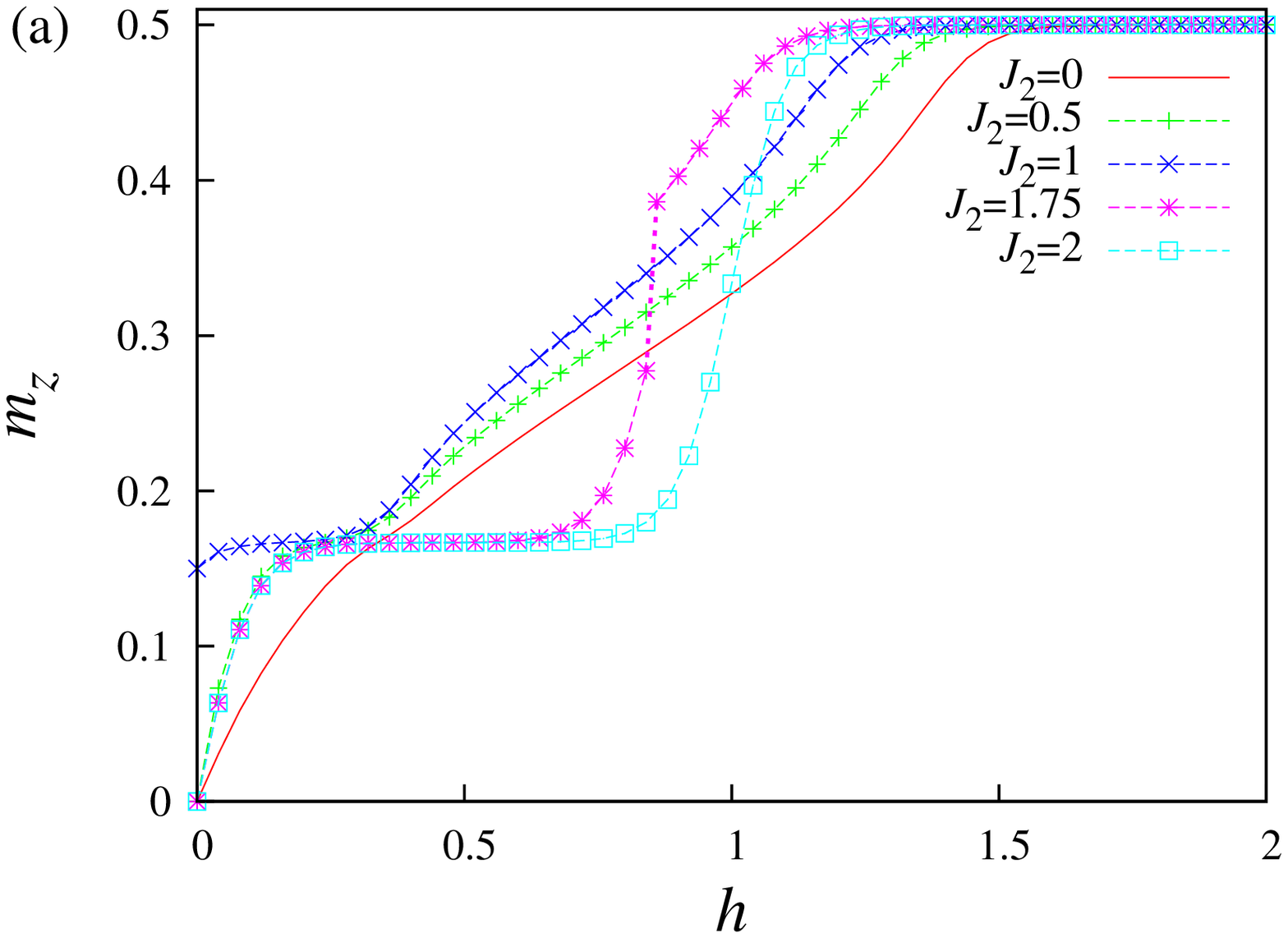}
}
\resizebox{0.9\columnwidth}{!}{%
  \includegraphics{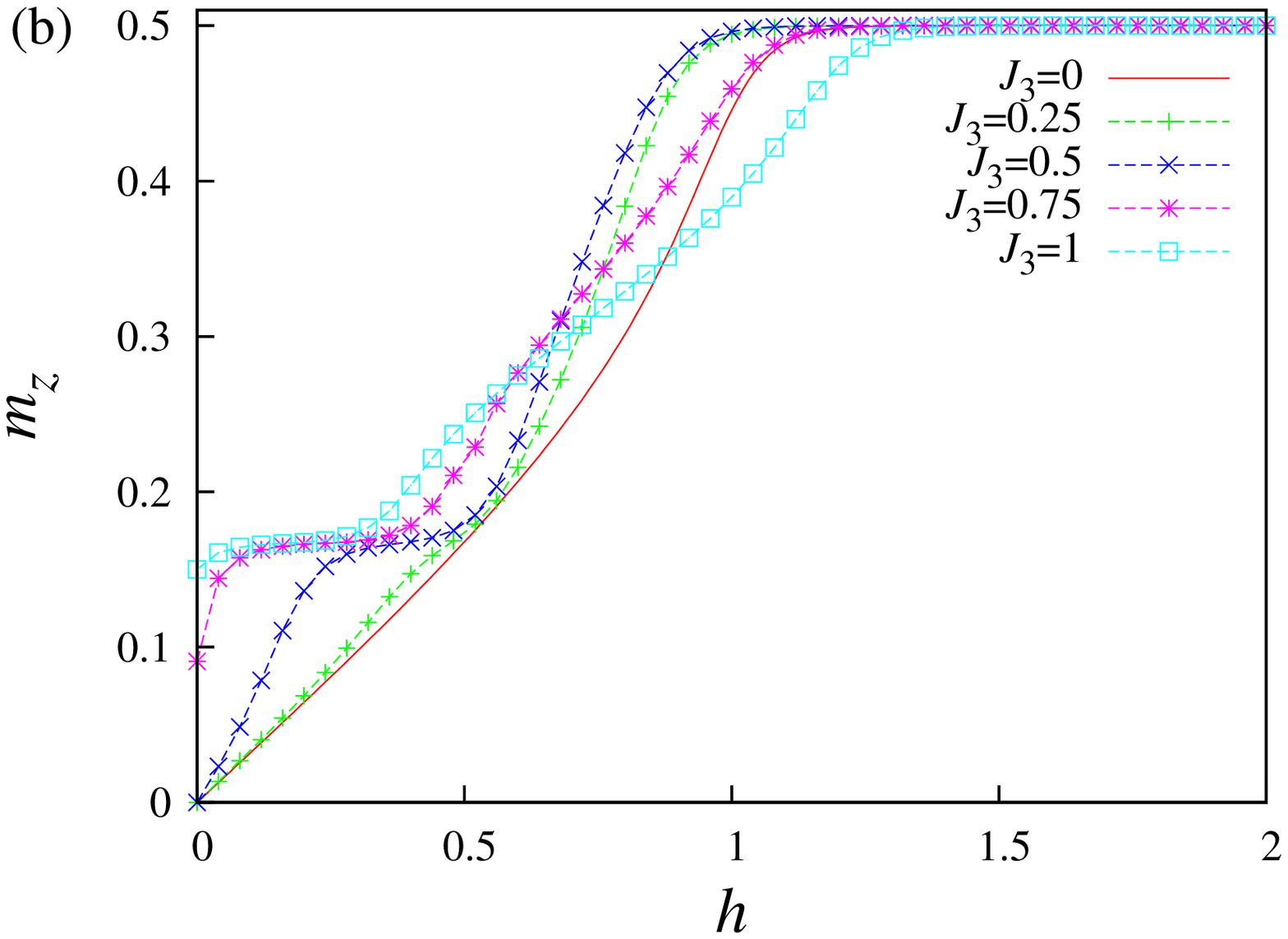}
}
\end{center}
\caption{(Color online) The magnetization in the Hartree-Fock approximation
vs. the external field of
the diamond chain for $T=0.05J_1$: a) $J_1=J_3=1$, $J_2=0, 0.5, 1, 1.75,2$;
b) $J_1=J_2=1$, $J_3=0, 0.25, 0.5, 0.75, 1$}
\label{fig_mag_t}
\end{figure}

The field dependent magnetization for non-zero temperatures
can be also understood.
The plateaux and jumps of the magnetization are smeared out by temperature
(see Fig.~\ref{fig_mag_t}).
One exception is the model when $J_2\sim J_1$ when the mean-field treatment preserves
jumps even for small temperatures.
Thus, the Hartree-Fock approximation turns out to be incorrect for
$J_1\sim J_2\sim J_3$ and small external field $h$, since the spontaneous (i.e.
non-zero) magnetization in the zero-field limit is prohibited for
one-dimensional system at any finite temperature due to the Mermin-Wagner
theorem \cite{mw}.

\section{Conclusions}
\label{conclusions}
In the present work, we consider the Jordan-Wigner transformation for the
spin-$\frac{1}{2}$ $XX$ model on the diamond chain by assuming a quite general
distorted case.

At first we showed that
the free-fermion model on the diamond chain cannot describe the spin system
since it loses  the symmetries of a spin model
and exhibits a non-zero magnetization in zero magnetic field.
In the case of the distorted chain we apply the Hartree-Fock approximation
for the fermionic representation of the model
which considers the fermion interaction along weaker bonds.
For the symmetric diamond chain we suggest the generalization of the
Jordan-Wigner transformation and build a fully symmetric fermionic representation
of the spin-model. Further, we use the Hartree-Fock approximation to
study the model near the limit of the dimer-monomer phase when the correlation
between the spins from the vertical bonds are the strongest one.
The Hartree-Fock approximation reproduces an exact result
for the dimer-monomer ground state
of the symmetric diamond chain ($J_2\geq 2J_1$).
Moreover, we have found that the solutions of the Hartree-Fock
approximation for the symmetric diamond chain are invariant
with respect to the gauge transformations.
Our results show also good agreement with the exact diagonalization data
reproducing the magnetic properties at high fields or small frustrations.
Summarizing our findings,
the elaborated approach for the $XX$ diamond chain
reproduces a plateau in the magnetization curve at 1/3 of the saturation
magnetization and an additional peak in the specific heat curve.
In fermionic language a 1/3 plateau is caused by the gap between two fermionic bands.
When the gap becomes large enough, the temperature dependence of the specific heat
gains a distinct two-peak structure.
Note that both features are typical for azurite
\cite{kikuchi2004,kikuchi2005a,kikuchi2005b,rule}
although the Heisenberg model seems to be more appropriate for that compound.

We have also observed some drawbacks of the Hartree-Fock treatment.
It becomes invalid for $J_1\sim J_2\sim J_3$.
In this case mean-fields leads to the non-zero magnetization in zero fields,
and the magnetization jump survives also for small temperatures.

\begin{acknowledgement}
The authors would like to thank O.~Derzhko for the useful discussions.
T.V. was supported by the National Scholarship Programme of the Slovak Republic.
J.S. and M.J. acknowledge the financial support by the ERDF EU
(European Union European regional development fund) grant provided under
the contract No. ITMS26220120005 (activity 3.2.).
\end{acknowledgement}

\appendix\section*{Appendix: Exact monomer-dimer ground state for the anisotropic diamond chain}

In case $J_1=J_3>0$ and $J_2\geq 2J_1$ the ground state of
the Hamiltonian on the diamond chain is the dimer-monomer state
\cite{tks}.
To prove this
we can follow the arguments of Shastry and Sutherland
to obtain the ground state \cite{shastry1981}.
We have to consider the Hamiltonian as a sum of Hamiltonians of triangles:
\begin{eqnarray}
H&=&\sum_l H_{l,l}+ H_{l,l+1},
\nonumber\\
H_{l,l}&=&\frac{J_2}{2}(\mathbf{s}_{2,l}\cdot\mathbf{s}_{3,l})_{\Delta}
+J_1\left((\mathbf{s}_{1,l}\cdot\mathbf{s}_{2,l})_{\Delta}
\right.
\nonumber\\
&&\left.
+(\mathbf{s}_{1,l}\cdot\mathbf{s}_{3,l})_{\Delta}\right)
-\frac{h}{2}\left(s_{1,l}^z+s_{2,l}^z+s_{3,l}^z\right),
\nonumber\\
H_{l,l+1}&=&\frac{J_2}{2}(\mathbf{s}_{2,l}\cdot\mathbf{s}_{3,l})_{\Delta}
+J_1\left((\mathbf{s}_{1,l+1}\cdot\mathbf{s}_{2,l})_{\Delta}
\right.
\nonumber\\
&&\left.
+(\mathbf{s}_{1,l+1}\cdot\mathbf{s}_{3,l})_{\Delta}\right)
-\frac{h}{2}\left(s_{1,l+1}^z+s_{2,l}^z+s_{3,l}^z\right).
\nonumber
\end{eqnarray}
Here we introduced the notation
$(\mathbf{s}_{p,l}\cdot\mathbf{s}_{q,m})_{\Delta}
=s_{p,l}^xs_{q,m}^x+s_{p,l}^ys_{q,m}^y+\Delta s_{p,l}^zs_{q,m}^z$.
The parts of the systems described by $H_{l,l}$ and $H_{l,l+1}$
are topologically identical.
Therefore, they have the same eigenvalues and analogous eigenstates.
One can use the variational principle that implies
$E_0\geq N(e_l+e_l)$, where $e_l$ is the ground state energy of
$H_{l,l}$ or $H_{l,l+1}$.

Let us consider the case $J_2>2J_1$.
The direct calculation shows that if
$|h|\leq h_c=\frac{\Delta(J_2+2J_1)+J_2}{2}$
the dimer-monomer state
$|\alpha_{1,l}\rangle[2l,3l]$
of $H_{l,l}$ or $H_{l,l+1}$ is the state with the lowest eigenvalue
$e_l^{dm}=-\frac{(\Delta+2)J_2}{8}-\frac{|h|}{4}$ and
$S_{tot}^z=\pm\frac{1}{2}$.
Here $|\alpha_{1,l}\rangle$ denotes the spin up state $|\uparrow_{1,l}\rangle$ if $h>0$,
or the spin down state $|\downarrow_{1,l}\rangle$ if $h<0$.
Thus using the variational principle the ground state of the whole crystal
can be built by translation of the state $|\alpha_{1,l}\rangle[2l,3l]$
over all sites. The total magnetization is $\frac{N}{2}\mbox{sgn}(h)$
and the magnetization per spin is $1/6$.
Contrary, if $|h|>h_c$, the completely polarized state
with $S_{tot}^z=\pm\frac{3}{2}$ becomes the state with the lowest eigenvalue
$e_l^{3/2}=\frac{\Delta(J_2+4 J_1)}{8}-\frac{3|h|}{4}$.
Thus, the total ground state is the state with all spins up or down,
with the magnetization per spin $m^z=1/2$.

To summarize, the field dependence of the ground-state magnetization
has a step-like form for $J_2>2J_1$, i.e., $m^z=\mbox{sgn}(h)/6$ if $|h|<h_c$
and $m^z=\mbox{sign}(h)/2$ otherwise,
whereas the critical field equals to
$h_c = J_1 \Delta + J_2 (1 + \Delta)/2$.


\begin{thebibliography}{99}

\bibitem{misguich} G. Misguich, C. Lhuillier,
   in: Frustrated Spin Systems, edited by H.T. Diep,
   (World Scientific, Singapore, 2004), p.229
\bibitem{mikeska} H.-J. Mikeska, A.K. Kolezhuk,
   in: Quantum Magnetism,
   edited by U. Schollw\"ock, J.~Richter, D.J.J.~Farnell, R.F.~Bishop,
   Lect. Notes Phys. {\bf 645}, (Springer, Berlin, 2004), p.1
\bibitem{richter} J. Richter, J. Schulenburg, A. Honecker,
   edited by U. Schollw\"ock, J.~Richter, D.J.J.~Farnell, R.F.~Bishop,
   Lect. Notes Phys. {\bf 645}, (Springer, Berlin, 2004), p.85

\bibitem{derzhko2007} O. Derzhko, J. Richter, A. Honecker, H.-J. Schmidt,
   Fizika Nizkikh Temperatur {\bf 33}, 982 (2007)
   [Low Temp. Phys. {\bf 33}, 745 (2007)]
\bibitem{prl02}
J. Schulenburg, A. Honecker, J. Schnack, J. Richter, H.-J.~Schmidt,
     Phys. Rev. Lett. {\bf 88}, 167207 (2002)

\bibitem{honecker04} A. Honecker,  J. Schulenburg, J. Richter,
            J. Phys.: Condens. Matter {\bf 16}, S749 (2004)

\bibitem{tks} K. Takano, K. Kubo, H. Sakamoto,
   J. Phys.: Condens. Matter {\bf 8}, 6405 (1996)

\bibitem{canova} L. \v{C}anov\'{a}, J. Stre\v{c}ka, M. Ja\v{s}\v{c}ur,
   J. Phys.: Condens. Matter {\bf 18}, 4967 (2006)

\bibitem{hida1} K. Takano, H. Suzuki, K. Hida,
   Phys. Rev. B {\bf 80}, 104410 (2009)

\bibitem{hida2} K. Hida, K. Takano, H. Suzuki,
   J. Phys. Soc. Jpn. {\bf 78}, 084716  (2009)

\bibitem{hida3} K. Hida, K. Takano, H. Suzuki,
   J. Phys. Soc. Jpn. {\bf 79}, 044702 (2010)

\bibitem{okamoto1999} K. Okamoto, T. Tonegawa, Yu. Takahashi, M. Kaburagi,
   J. Phys.: Condens. Matter {\bf 11}, 10485 (1999)
\bibitem{okamoto2003} K. Okamoto, T. Tonegawa, M. Kaburagi,
   J. Phys.: Condens. Matter {\bf 15}, 5979 (2003)

\bibitem{okamoto2005}
K. Okamoto, A. Tokuno, Y. Ichikawa,
   J. Phys. Chem. Solids {\bf 66}, 1442 (2005)

\bibitem{okamoto2005-2}
A. Tokuno, K. Okamoto,
   J. Phys. Soc. Jpn. Suppl. {\bf 74}, 157  (2005)

\bibitem{okamoto2007}
K. Okamoto, A. Tokuno, T. Sakai,
   J. Magn. Magn. Mater., {\bf 310}, e457-e459 (2007)

\bibitem{kikuchi} H. Kikuchi, Y. Fujii, M. Chiba, S. Mitsudo, T. Idehara,
   Physica B {\bf 329-333} 967 (2003)

\bibitem{hosokoshi} Y. Hosokoshi, K. Katoh, Y. Nakazawa, H. Nakano, K. Inoue,
   J. Am. Chem. Soc. {\bf 123}, 7921 (2001)

\bibitem{sakurai}
   H. Sakurai, K. Yoshimura, K. Kosuge
   N.~Tsujii, H.~Abe, H.~Kitazawa, G.~Kido, H.~Michor, G.~Hilscher,
   J. Phys. Soc. Jpn. \textbf{71}, 1161 (2002)

\bibitem{uematsu}
D. Uematsu, M. Sato, J. Phys. Soc. Jpn. \textbf{76}, 084712 (2007)

\bibitem{lazari}
G. Lazari, T. C. Stamatatos, C. P. Raptopoulou
V.~Psycharis, M.~Pissas, S.P.~Perlepes, A.K.~Boudalis,
Dalton Trans. 3215 (2009)

\bibitem{ohta2003} H. Ohta, S. Okubo, T. Kamikawa \textit{et al}.,
   J. Phys. Soc. Jpn. \textbf{72}, 2464 (2003)

\bibitem{kikuchi2004} H. Kikuchi, Y. Fujii, M. Chiba \textit{et al}.,
   J. Magn. Magn. Mater. \textbf{272-276}, 900 (2004)

\bibitem{ohta2004} H. Ohta, S. Okubo, Y. Inagaki \textit{et al}.,
   Physica~B \textbf{346-347}, 38 (2004)

\bibitem{kikuchi2005a} H. Kikuchi, Y. Fujii, M. Chiba \textit{et al}.,
   Progr. Theor. Phys. Suppl. \textbf{159}, 1 (2005)

\bibitem{kikuchi2005b} H. Kikuchi, Y. Fujii, M. Chiba, S. Mitsudo, T. Idehara, T. Tonegawa,
   K. Okamoto, T. Sakai, T. Kuwai, H. Ohta,
   Phys. Rev. Lett. {\bf 94}, 227201 (2005)

\bibitem{rule}
K.C. Rule, A.U.B. Wolter, S. Sullow, D.A. Tennant, A. Bruhl, S. Kohler,
   B. Wolf, M. Lang, J. Schreuer,
   Phys. Rev. Lett. {\bf 100}, 117202 (2008)

\bibitem{mikeska2008}  H.-J. Mikeska, C. Luckmann,
   Phys. Rev. B {\bf 77}, 054405 (2008)

\bibitem{fu} H. H. Fu, K. L. Yao, Z. L. Liu,
   Phys. Rev. B {\bf 73}, 104454 (2006);
   Phys. Rev. B {\bf 77}, 219901(E) (2008)

\bibitem{gu} B. Gu, G. Su,
   Phys. Rev. B {\bf 75}, 174437 (2007)

\bibitem{azurite2010} H. Jeschke, I. Opahle, H. Kandpal, \textit{et al}.,
   arXiv:1012.1090v1 [cond-mat.str-el]

\bibitem{azzouz1994} M. Azzouz, Liang Chen, S. Moukouri,
   Phys. Rev. B {\bf 50}, 6233 (1994)

\bibitem{nunner} T. S. Nunner, Th. Kopp,
   Phys. Rev. B {\bf 69}, 104419 (2004)

\bibitem{verkholyak2006} T. Verkholyak, A. Honecker, W. Brenig,
   Eur. Phys. J. B. {\bf 49}, 283 (2006)

\bibitem{mueller1981} G. M\"uller, H. Thomas, H. Beck, J. C. Bonner,
   Phys. Rev. B {\bf 24}, 1429 (1981)

\bibitem{derzhko2005} O. Derzhko, T. Krokhmalskii, J. Stolze, G. M\"uller,
   Phys. Rev. B {\bf 71}, 104432 (2005)

\bibitem{lsm} E.~Lieb, T.~Schultz, D.~Mattis,
   Ann. Phys. (N.Y.) {\bf 16}, 407 (1961)

\bibitem{katsura} S.~Katsura,
   Phys. Rev. {\bf 127}, 1508 (1962);
   {\bf 129}, 2835 (1963)

\bibitem{zaspel1987} C. E. Zaspel,
    J. Chem. Phys. {\bf 86}, 4713 (1987)

\bibitem{okamoto1992} K. Okamoto,
   Solid State Commun.  {\bf 83}, 1039 (1992)

\bibitem{okamoto1996} K. Okamoto,
   Solid State Commun.  {\bf 98}, 245 (1996)

\bibitem{derzhko1999} O. Derzhko, J. Richter, V. Derzhko,
   Ann.Phys. (Leipzig) {\bf 8}, SI-49 (1999); cond-mat/9908425v1

\bibitem{elk} K. Elk, W. Gasser,
   Die Methode der Greenschen Funktionen in der Festk\"orperphysik,
   (Akademie-Verlag, Berlin, 1979)

\bibitem{dmitriev2002} D.V. Dmitriev, V.Y. Krivnov, A.A. Ovchinnikov,
   Phys. Rev. B {\bf 65}, 172409 (2002)

\bibitem{caux2003} J.-S. Caux, F. H. L. Essler, U. L\"ow,
   Phys. Rev. B {\bf 68}, 134431 (2003)
\bibitem{hagemans2005}   R. Hagemans, J.-S. Caux, U. L\"ow,
   Phys. Rev. B {\bf 71}, 014437 (2005)

\bibitem{shastry1981} B. S. Shastry, B. Sutherland, Physica B {\bf 108}, 1069 (1981);
   Phys. Rev. Lett. {\bf 47}, 964 (1981)

\bibitem{li2007} Yan-Chao Li,
J. Appl. Phys. {\bf 102}, 113907 (2007)

\bibitem{mw} N. D. Mermin, H. Wagner, Phys. Rev. Lett. {\bf 17}, 1133 (1966)

\end{thebibliography}
\end{document}